\documentclass[a4paper]{article}
\usepackage[english]{babel}
\usepackage[pdftex]{graphicx}
\usepackage[latin1]{inputenc}
\usepackage{amsmath,amsthm,amsfonts,eucal,mathrsfs}
\usepackage{a4wide}

\newcommand{\lint}{\int\limits}
\newcommand{\e}{\mathbf{e}}
\newcommand{\E}{\CMcal{E}}
\newcommand{\R}{\mathbb{R}}
\newcommand{\T}{\CMcal{T}}
\newcommand{\vv}{\mathbf{v}}
\newcommand{\x}{\mathbf{x}}
\newcommand{\X}{\mathbf{X}}
\newcommand{\xibo}{\boldsymbol\xi}
\DeclareMathOperator*{\argmax}{arg\,max}

\graphicspath{{./}{figure/figure_arXiv/}}

\long\def\symbolfootnote[#1]#2{\begingroup%
\def\thefootnote{\fnsymbol{footnote}}\footnote[#1]{#2}\endgroup} 

\begin{document}

\title{Non-local first-order modelling of crowd dynamics:\\ a multidimensional framework with applications}

\author{Luca Bruno\textsuperscript{a},
		Andrea Tosin\textsuperscript{b}\footnote{Corresponding author. Tel: (+39) 011.090.7531.\newline
			{\it E-mail address}: \texttt{andrea.tosin@polito.it} (Andrea Tosin).},
		Paolo Tricerri\textsuperscript{c}, 
		Fiammetta Venuti\textsuperscript{a}\\[0.5cm]
	{\small\it\textsuperscript{a}Politecnico di Torino, Department of Structural and Geotechnical 
		Engineering,}\\[-0.1cm]
	{\small\it Viale Mattioli 39, I-10125, Torino, Italy}\\
	{\small\it\textsuperscript{b}Politecnico di Torino, Department of Mathematics,}\\[-0.1cm]
	{\small\it Corso Duca degli Abruzzi 24, I-10129, Torino, Italy}\\
	{\small\it\textsuperscript{c}Ecole Polytechnique F\'ed\'erale de Lausanne,}\\[-0.1cm]
	{\small\it Mathematics Institute of Computational Science and Engineering,}\\[-0.1cm]
	{\small\it Station 8, CH-1015, Lausanne, Switzerland}
}
		
\date{}

\maketitle

\begin{abstract}
In this work a physical modelling framework is presented, describing the intelligent, non-local, and anisotropic behaviour of pedestrians. Its phenomenological basics and constitutive elements are detailed, and a qualitative analysis is provided. Within this common framework, two first-order mathematical models, along with related numerical solution techniques, are derived. The models are oriented to specific real world applications: a one-dimensional model of crowd-structure interaction in footbridges and a two-dimensional model of pedestrian flow in an underground station with several obstacles and exits. The noticeable heterogeneity of the applications demonstrates the significance of the physical framework and its versatility in addressing different engineering problems. The results of the simulations point out the key role played by the physiological and psychological features of human perception on the overall crowd dynamics.
\end{abstract}

%
%
%
%

\section{Introduction}
\label{one}
In relatively recent years, crowd dynamics has stimulated the interest of scientists such as physicists, engineers, and applied mathematicians. This has been mainly caused by some media events, sometimes catastrophic, which have pointed out the necessity to complement traditional methods of scientific investigation, heavily relying on empirical observations, with prediction and simulation tools. For instance, in Civil Engineering it is well-known the case of the London Millennium Bridge, closed immediately after its opening because of excessive lateral oscillations developing when pedestrians crossed it. The bridge was reopened two years later after the installation of dampers to avoid uncomfortable vibrations. The intense research activity that has followed the event up to now (reviewed e.g., in \cite{ziv,venPLR}) found that those oscillations were due to unpredicted resonances triggered by synchronised walkers \cite{dal} and showed the importance to consider also the crowd-structure coupling when designing pedestrian facilities. Another significant example is the Jamarat Bridge in Mina (South Arabia), annual destination of Muslim pilgrims. The huge amount of people cramming the bridge produced in the Nineties hundreds of deaths for overcompression. In this case, mathematical models were successfully employed to simulate the flow of pedestrians and to investigate its relationships with the arrangement of the surrounding environment. This made it possible to highlight critical situations in which crowd disasters tended to occur and suggest effective countermeasures to improve the safety of the event \cite{helb_5}. Aside from these serious incidents, crowded places can be commonly experienced in daily life: shopping centres, railway or underground stations, airports, stadia, whose design may take advantage of virtual simulations to preliminary study and test different solutions, also in connection with safety and optimisation issues. This requires to set up suitable models of pedestrian behaviour, on one side accurate to catch the complex unsteady crowd dynamics and, on the other side, versatile and easy to handle to deal with real world applications. 

The modelling of pedestrian flows dates back at least to the Sixties, when experimental studies started to be flanked by \emph{ad-hoc} models aiming at reproducing specific aspects of pedestrian behaviour, such as queuing and route choice, see e.g., \cite{ceder1979aap} and the review \cite{helbing2009pce}. The first model suited to reproduce spatio-temporal patterns of pedestrian movement is probably \cite{hen}, proposed in the first half of the Seventies. In recent years, many mathematical models have been proposed in the literature to describe crowd dynamics, at both the microscopic \cite{helb_2,HeMoFaBo,HoBo_ped,maury2008mfc,MR2556188,singh20094408} and the macroscopic \cite{bel4,col3,cos,hug,maury2009mcm-preprint,piccoli2009pfb} scale. The physiological and psychological features of human perception are expected to play a key role in pedestrian behaviour. In particular, the visual perception implies that pedestrians acquire and make synthesis of non-local information about their surroundings, as widely investigated in the medical \cite{patla} and cognitive psychology \cite{gib,koss} fields. The non-local pedestrian perception suggests the need for developing non-local crowd models. Some features of visual perception (e.g., the visual field in the horizontal plane) have been considered in developing visually guided agent-based (microscopic) models of pedestrian behaviour in transportation engineering (see e.g., \cite{robi} and referenced papers) and in architecture at both building and urban scales (see e.g., \cite{tur}). A hint toward the use of space non-locality in first-order (i.e., mass-conservation-equation-based) macroscopic models can be found in \cite{bel3} for vehicular traffic applications, in order to allow for instability phenomena characteristic of high density flows. Concerning pedestrians, some proposals to account for their anisotropic and non-local behaviour at the macroscopic scale are contained in recently proposed models \cite{bel4,piccoli2009pfb}.

In this paper a phenomenological analysis is performed to point out the intelligent, non-local, and anisotropic features of pedestrian behaviour in normal (no panic) conditions. On these bases, a physical modelling framework is proposed. Then, two first-order mathematical models are derived, oriented to specific real world applications: a one-dimensional (1D) model to simulate crowd-structure interaction in footbridges and a two-dimensional (2D) model to simulate the crowd flow in an underground station with complex plan. These models resort to non-linear mass conservation laws with non-local fluxes, hence they directly deduce pedestrians' velocity from phenomenological principles, and ultimately provide a macroscopic description of the spatio-temporal crowd dynamics. The paper is organised into three more sections that follow the conceptual approach described above. In Sect. \ref{two}, the phenomenological analysis, the physical model, and its qualitative analysis are described. In Sect. \ref{three}, two different mathematical models and the related numerical solution techniques are derived to handle 1D and 2D crowd dynamics for real world applications. Finally, in Sect. \ref{four} conclusions are outlined and some further research perspectives are briefly sketched.

\section{Physical framework}
\label{two}

\subsection{Phenomenological analysis}
\label{two-1}
The proposed physical modelling framework is developed on the basis of a phenomenological analysis of crowd dynamics, specially focussed on the role played by pedestrian perception, evaluation, and reaction to the surrounding conditions. In the following, the retained features of pedestrian behaviour are summarised in three conceptual blocks.

Remarks about \emph{intelligent behaviour}:
\begin{enumerate}
\item[(f1)] pedestrians are \emph{active} agents, i.e., under normal conditions without panic onset, they share the same objective of walking with the maximum velocity towards a target $\T$ (e.g., doors, exit, displays), bypassing possible obstacles and avoiding the most crowded zones (see the concept of \emph{personal space} developed in human sciences and psychology, e.g. in \cite{dos}). These strategies enable them to determine actively their walking direction and velocity, without being passively subject to the laws of inertia;

\item[(f2)] pedestrians are \emph{intelligent} agents, i.e., their mind evaluates, selects, and/or makes synthesis of what it perceives according to various psychological criteria (e.g., the level of anxiety \cite{bra} or the capacity to perform ensamble evaluations \cite{koss}). 
\end{enumerate}

Remarks about \emph{non-local behaviour}:
\begin{enumerate}
\item[(f3)] under normal external and subjective conditions (e.g., the area where pedestrians walk, hereafter called \emph{walking area}, is illuminated allowing visual perception), pedestrians do not perceive the real world locally in space, due to their ability to see up to a given extent around them. Let us call \emph{sensory region} ``the area required by pedestrians for perception, evaluation and reaction'', previously introduced by Fruin in his pioneering work \cite{fruin}, and let us notate it $R_s$;

\item[(f4)] pedestrians react after a time interval has elapsed from the perception time (see e.g., \cite{patla,koss}): from a general point of view, a \emph{reflex time delay} $\tau_1$, which does not involve a long perception time nor synthesis, is needed for reaction to the actual conditions, while a \emph{volitional time delay} $\tau_2$ is required to scan the visual field and to adopt walking strategies. It is expected that $\tau_1\ll\tau_2\ll T$, $T$ being the reference macroscopic time scale required to cross the walking area at the maximum walking speed $v_M$;

\item[(f5)] from the previous features, it comes that pedestrians in a given position at a given time react to the conditions perceived in front of them at a delayed time, i.e., in a \emph{non-local} way in both space and time.
\end{enumerate}

Remarks about \emph{anisotropic behaviour}:
\begin{enumerate}
\item[(f6)] pedestrians are \emph{anisotropic} agents, i.e., they are not equally affected by stimuli coming from all directions in space. Specifically, they distinguish between ahead and behind, in normal situations being essentially sensitive to what happens in a symmetric visual field focused on their direction of movement;

\item[(f7)] walking pedestrians adapt the depth and width of the sensory region to their travel purpose and to their walking speed (e.g., a pedestrian walking for leisure purposes is expected to scan a wider field than a commuter attaining a train, and the faster a pedestrian walks the deeper the space required to evaluate and react is).
\end{enumerate}

\subsection{Physical model}
\label{two-2}
\begin{figure}[!t]
\centering
\includegraphics{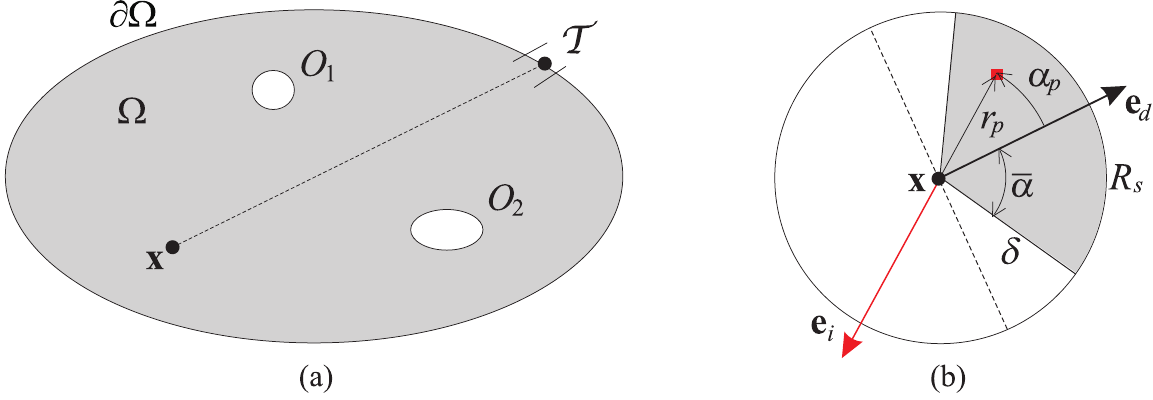}
\caption{(a) Walking area $\Omega$ with obstacles ($\x$=pedestrian position, $\T$=target); (b) Sensory region $R_s$, desired direction $\e_d$, and interaction direction $\e_i$}
\label{fig.1}
\end{figure}

On the basis of the above mentioned features, a physical modelling framework is proposed. Let $\Omega$ be a bounded domain representing the walking area (Fig. \ref{fig.1}a), and assume that the target $\T$ lies on the boundary $\partial\Omega$. The proposed model is expressed by a mass conservation equation, herein written for a control mass $M$:
\begin{equation}
\dfrac{DM}{Dt}=0,
\label{Lf-mc}
\end{equation}
where $t$ is time.

In order to fully describe the dynamics of the system, the velocity field of the moving mass should be determined as a state variable or be \emph{a priori} known. First-order models relate the velocity field to the mass itself.

Let us introduce the pedestrian velocity field $\vv$ in the point $\x\in\Omega$ at time $t$:
\begin{equation}
\vv(\x,t)=v(\x,t)\e_v(\x,t),
\label{eq.v}
\end{equation}
where $v$ is the magnitude of $\vv$ (i.e., the walking speed) and $\e_v$ is a unit vector representing the walking direction. In order to account for the contribution of the strategies outlined in point f1, the walking direction is modelled as the superposition of two contributions:

\begin{enumerate}
\item[(m1)] the \emph{desired direction}, i.e., the direction that pedestrians would follow to reach their target in a clear domain (the so-called unconstrained conditions \cite{daa}). It is basically affected by the geometrical configuration of the boundaries of the walking area e.g., perimeter walls and/or inner obstacles, and it does not evolve in time. This direction is identified by the unit vector $\e_d$ (Fig. \ref{fig.1}b);

\item[(m2)] the \emph{interaction direction}, i.e., the direction that pedestrians would follow to avoid crowded zones. This direction is induced by the interactions among pedestrians and is identified by the unit vector $\e_i$ (Fig. \ref{fig.1}b), which, unlike $\e_d$, evolves in time.
\end{enumerate}
It is worth pointing out that, in general, both directions may be the result of a perception process. In particular, both the geometry of the walking area and the crowding of the environment might be discovered by pedestrians during walking. Nevertheless, in the following pedestrians are assumed to already know the entire walking area from previous experiences, so that the desired direction $\e_d$ is \emph{a priori} determined. The knowledge of crowding and the related interaction direction are instead obtained through a perception process characterised by a non-local and anisotropic behaviour.

The resulting walking direction $\e_v$ is expressed as:
\begin{equation}
\e_v=\frac{\theta\e_d+(1-\theta)\e_i}{\vert\theta\e_d+(1-\theta)\e_i\vert},
\label{eq.dir}
\end{equation}
where $\theta\in[0,\,1]$ is a dimensionless parameter that weights pedestrian attitude to give priority to the walking area layout or to the crowd conditions. It is worth noticing that reversed pedestrian flow occurs if $\theta\in[0,\,0.5]$. Notice that in the one-dimensional setting Eq. \eqref{eq.dir} simply reduces to $\e_v=\mathbf{i}$, where $\mathbf{i}$ is the unit vector in the direction of the horizontal axis, because the dimensional constraint forces both $\e_d$ and $\e_i$ to be directed along $\mathbf{i}$.

\begin{figure}[!t]
\centering
\includegraphics{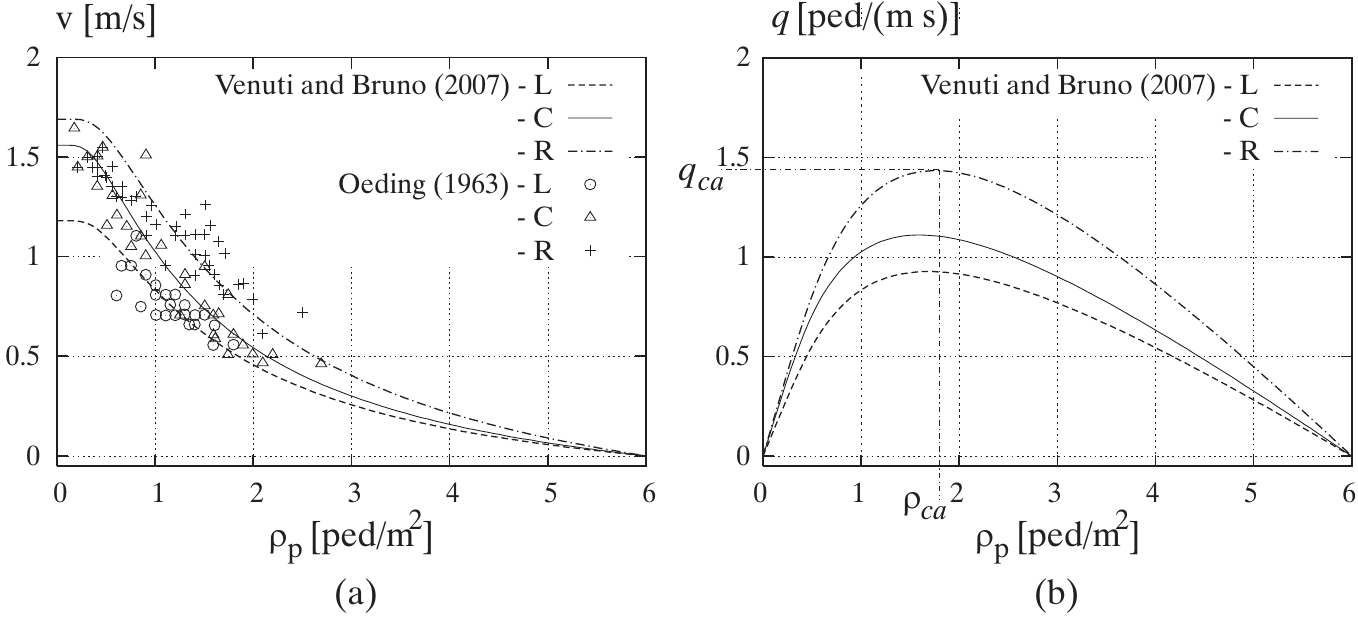}
\caption{(a) Speed-density laws as proposed in \cite{venCRM1} and (b) related flow-density diagrams}
\label{fig.2}
\end{figure}

The walking speed $v$ is generally determined as a function of the crowd density through one of the constitutive laws proposed in literature (the so-called fundamental relations reviewed in \cite{daa} or recently proposed in \cite{helb_5,col3}). In the following, the Kladek formula introduced by Weidmann \cite{weid} is adopted in the revisited version proposed in \cite{venCRM1}: 
\begin{equation}
v=v_M\left\{1-\exp{\left[-\gamma\left(\frac{1}{\rho_p}-\frac{1}{\rho_M}\right)\right]}\right\},
\label{eq.kla}
\end{equation}
where $v_M$ is the maximum walking speed in unconstrained conditions (also called free speed); $\rho_M$ is the jam density; $\gamma$ is an exponent, obtained through a fitting of the data in \cite{fruin,oed} (Fig. \ref{fig.2}), that makes the relation sensitive to different travel purposes (L = leisure/shopping, C = commuters/events, R = rush hour/business); $\rho_p=\rho_p(\x,\,t)$ is the \emph{perceived density} that a pedestrian located in $\x$ feels within his/her sensory region.

Both magnitude and direction of the walking velocity require the modelling of the non-local anisotropic behaviour of pedestrians (points f3 to f7).

The sensory region $R_s=R_s(\x,\,\e_d,\,\delta,\,\bar{\alpha})$ is modelled as a portion of the ball centred in $\x$ with radius $\delta$ and angular span $2 \bar{\alpha}$, which reproduce the characteristic depth and width of the visual field of pedestrians, respectively (Fig. \ref{fig.1}b, point f7).
Hence, in a 2D domain the sensory region is defined as:
\begin{subequations}
\begin{equation}
R_s(\x,\,\e_d,\,\delta,\,\bar{\alpha})=\left\{\xibo\in\Omega:\vert\xibo-\x\vert\leq\delta,\ 
	\frac{\xibo-\x}{\vert\xibo-\x\vert}\cdot\e_d(\x)\geq\cos{\bar{\alpha}}\right\},
\label{eq.Rs.2D}
\end{equation}
In the 1D case this definition simplifies as
\begin{equation}
R_s(x,\,\delta)=[x,\,x+\delta]\subseteq[0,\,L],
\label{eq.Rs.1D}
\end{equation}
\end{subequations}
the interval $\Omega=[0,\,L]$ being the 1D domain. The visual field is oriented according to the desired direction $\e_d$ and is symmetric with respect to the latter, because of pedestrian attitude to bear in mind their target (point f6). Within the perspective of a simplified model, let us assume that the depth $\delta$ depends on the walking speed only, while the width $2\bar{\alpha}$ is assumed to be constant. In particular, the radius $\delta$ is expressed as a function of the local delayed walking speed:
\begin{equation}
\delta(\x,\,t)=\delta(v(\x,\,t-\tau_1)),
\label{eq.delta}
\end{equation}
where $\tau_1$ is the reflex time delay introduced in point f4. It is worth noticing that $\delta$ should be more properly computed with reference to the walking speed in the point $\x'$, namely the position at time $t-\tau_1$ of the agent who at time $t$ is in $\x$, rather than in $\x$ itself. This essentially amounts to track pedestrian trajectories under a Lagrangian point of view, whereas Eq. \eqref{eq.delta} tries to keep an Eulerian reference. Such an approximation is however acceptable in view of the usual smallness of the reflex time delay $\tau_1$, which, from the mathematical point of view, implies a small error $\vert\x'-\x\vert$ for a smooth velocity field $\vv$.

\begin{figure}[!t]
\centering
\includegraphics{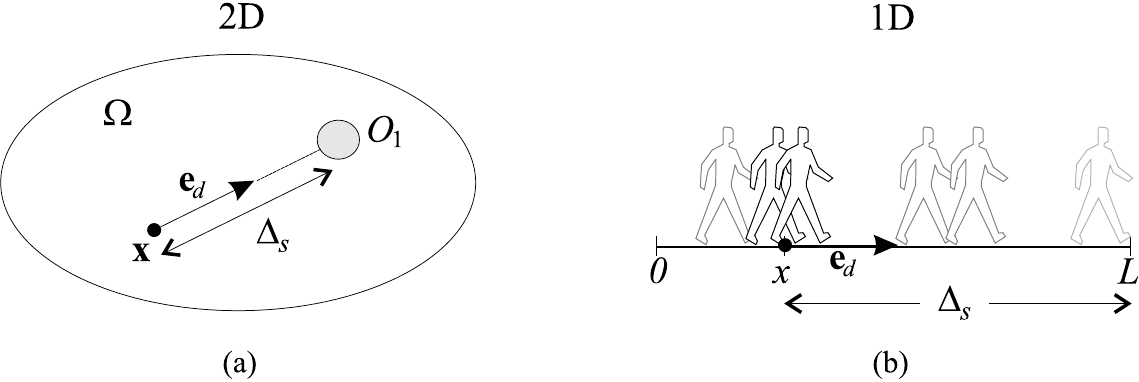}
\caption{Examples of $\Delta_s$ in 2D and 1D spatial domains}
\label{fig.3}
\end{figure}

\begin{figure}[!t]
\centering
\includegraphics{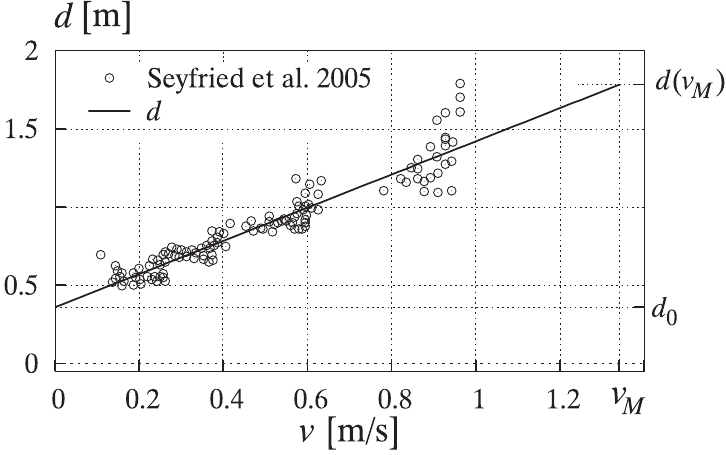}
\caption{$d-v$ law proposed in \cite{sey}}
\label{fig.4}
\end{figure}

The constitutive law linking $\delta$ to the walking speed $v$ is expressed as follows:
\begin{equation}
\delta(v)=\frac{\Delta_s}{v_M}v+\delta_0,
\label{eq.delta_v}
\end{equation}
where $\Delta_s=\Delta_s(\x)$ and $\delta_0>0$ correspond to the maximum and minimum visual depth, respectively, along the direction $\e_d$ in the empty walking area (Fig. \ref{fig.3}).
The expression of $\delta(v)$ is inspired to the distance $d$ of two successive pedestrians, defined at the microscopic scale in \cite{sey} on the basis of experimental results (Fig. \ref{fig.4}). In this perspective, $\Delta_s$ and $\delta_0$ are understood as the macroscopic counterparts of $d(v_M)$ and $d_0$, respectively. In particular, $\delta_0$ is assumed to be a constant equalling a given fraction of the reference length $L$.

The visual field width $\bar{\alpha}$ varies in the range $[0,\,90^\circ]$: in the following, the value $\bar{\alpha}=85^{\circ}$ is adopted according to \cite{robi}. It is worth stressing that the visual focus is higher in the neighbourhood of $\alpha=0$ and decreases for $\alpha$ approaching $\bar{\alpha}$ \cite{gib}.

The crowd density $\rho_p=\rho_p(\x,\,t)$ and, in the 2D setting, also the interaction direction $\e_i$ come from an intelligent evaluation process (points f2 and f6), which implies some weighting of the true density $\rho$. Different evaluation processes can be modelled, in order to describe the main features of the intelligent human behaviour within a crowd. In the sequel, four weighting strategies s1, \dots, s4 will be presented, leading to different definitions of $\rho_p$ and $\e_i$ in terms of $\rho$ in $R_s$. All of them can be formulated following a common conceptual line:

\begin{enumerate}
\item[(i)] \emph{Localisation}. A point $\x_p\in R_s$ is localised by means of a functional relationship in space involving the true density $\rho$:
$$ \x_p=\Psi[\rho], $$
where $\Psi$ is some localisation operator in $R_s$. The point $\x_p$ has to be understood as the location where crowding is influential according to the specific perception criterion expressed by $\Psi$.

\item[(ii)] \emph{Perception of the density}. In the strategies s1 and s2, the point $\x_p$ is used to evaluate the density $\rho_p$ perceived in $\x$ as
$$ \rho_p(\x,\,t)=\rho(\x_p,\,t), $$
meaning that pedestrians are mainly affected by the crowd in the point that has drawn their attention. In the strategies s3 and s4, the perceived density is the result of an ensemble evaluation of the crowd distribution over a part of or in the whole sensory region.

\item[(iii)] \emph{Route choice}. Once $\rho_p$ is known, the walking speed $v$ is computed from the fundamental relation \eqref{eq.kla}. Furthermore, in the two-dimensional case, the interaction direction $\e_i$ is recovered from $\x_p$ as
\begin{equation}
	\e_i(\x,\,t)=-\frac{\x_p-\x}{\vert\x_p-\x\vert},
	\label{eq.ei}
\end{equation}
the minus sign being due to that interactions lead pedestrians to steer clear of the crowd perceived in $\x_p$.
\end{enumerate}

Four possible definitions of the localisation operator $\Psi$, amounting to as many behavioural strategies from pedestrians, are now discussed, pointing out different psychological attitudes.
\begin{itemize}
\item[(s1)] Pedestrians evaluate the perceived density at the intersection between the far boundary of the sensory region and the desired direction, independently of the density distribution:
\begin{subequations}
\begin{equation}
	\x_p=\x+\delta\e_d(\x),
	\label{eq.xp.s1.2D}
\end{equation}
which in the 1D case reduces to fixing
\begin{equation}
	x_p=x+\delta.
	\label{eq.xp.s1.1D}
\end{equation}
\end{subequations}

In this strategy, pedestrians are expected to be strongly determined to reach an attracting goal (e.g., pedestrians leaving a room with one exit), so that they are slightly interested in the surrounding conditions. Hence, an \emph{a priori} localisation process is adopted, focussing on the far boundary of the sensory region along the desired direction.

\item[(s2)] Pedestrian attention is drawn by the point in the sensory region where the maximum value of $\rho$ is attained (in case of multiple points with the same maximal density, the closest one to $\x$ is chosen):
\begin{equation}
	\x_p=\Psi[\rho]=\argmax_{\xibo\in R_s}\rho(\xibo,\,t).
	\label{eq.xp_argmax}
\end{equation}
An analogous expression holds in the 1D case, up to maximising $\rho(\xi,\,t)$ for $\xi\in[x,\,x+\delta]$.

This strategy is intended to account for pedestrian anxiety about high crowd density (e.g., pedestrians leaving a room with multiple exits), hence an \emph{a posteriori} localisation is adopted, based on the observation of the environmental conditions. However, a pointwise criterion is used because pedestrian anxiety does not allow for any kind of ensemble evaluation of the crowd conditions.

\item[(s3)] The point $\x_p$ is still given by Eq. \eqref{eq.xp_argmax}, but the perceived density is a weighted average of the true density in $\x$ and in $\x_p$ based on the distance $r_p=\vert\x_p-\x\vert$:
\begin{equation}
\rho_p(\x,\,t)=(1-g(r_p))\rho(\x,\,t)+g(r_p)\rho(\x_p,\,t),
\label{eq.rhop.s3}
\end{equation}
where
$$ g(r_p)=-\frac{8}{10\delta}r_p+1. $$
Since $0\leq r_p\leq\delta$, the function $g$ ranges in $[0.2,\,1]$ and is monotonically decreasing with $r_p$. In particular, the maximum $g(0)=1$ reflects the fact that a pedestrian located in the point of maximal crowd density gives it the greatest importance, so as to avoid contact with other people. The form of Eq. \eqref{eq.rhop.s3} in the 1D case is straightforward.

In this strategy, pedestrians are still mainly concerned with the highest crowd density, but they are able to give it due weight with respect to its relative position. Hence, the evaluation of the perceived density implies self-consciousness from pedestrians, who establish a relation between themselves and the environmental conditions (e.g., commuters attaining their final destination in rush hours).

\begin{figure}[!t]
\centering
\includegraphics[scale=0.6,clip]{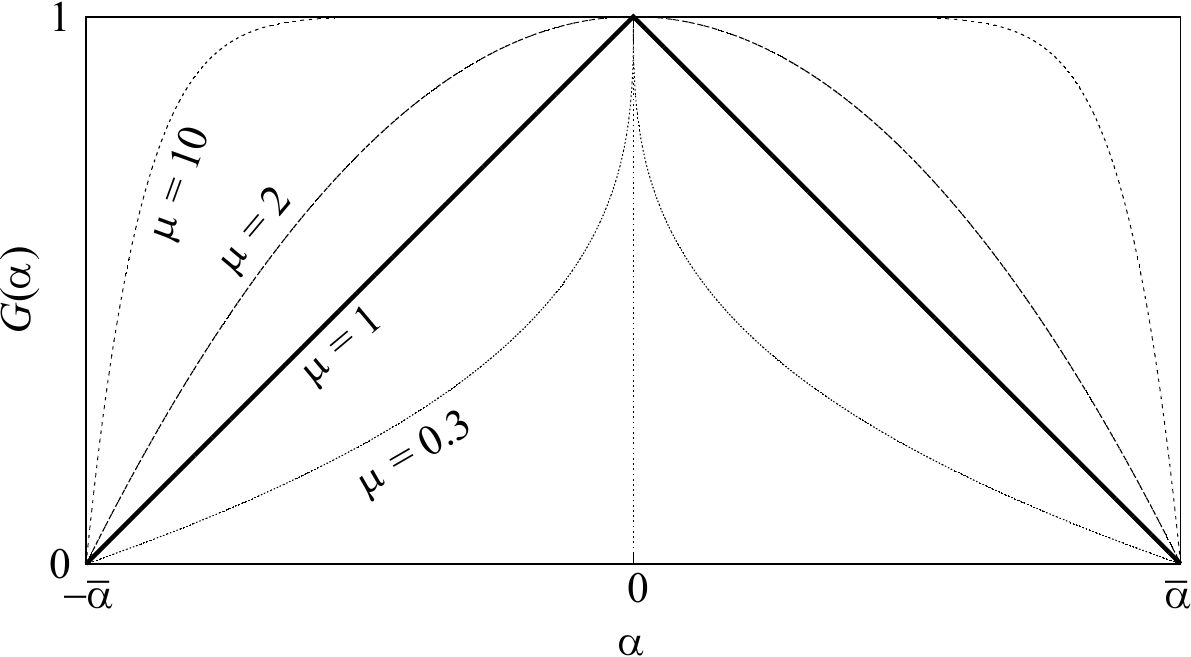}
\caption{The family of functions \eqref{eq.G.s4}}
\label{fig.5}
\end{figure}

\item[(s4)] Pedestrians evaluate $\x_p$ as the centre of mass of the whole crowd distributed in $R_s$, and use it as an indicator of the location of the mean value of the crowd density. However, when scanning the angular width of the sensory region they focus mainly on the frontal zone (cf. also \cite{gib}), hence the detection of the centre of mass is weighted by the angular position $-\bar{\alpha}\leq\alpha\leq\bar{\alpha}$ in $R_s$, see Fig. \ref{fig.1}b. Weighting is carried out by a function $G:[-\bar{\alpha},\,\bar{\alpha}]\to[0,\,1]$, which, coherently with the symmetry of $R_s$ around $\e_d$, is assumed to be even and monotonically decreasing from $G(0)=1$ to $G(\bar{\alpha})=0$ to render the lateral visual fading. Among different classes of weighting functions complying with these criteria, the following is considered:
\begin{subequations}
\begin{equation}
	G(\alpha)=1-\left(\frac{\vert\alpha\vert}{\bar{\alpha}}\right)^\mu,
	\label{eq.G.s4}
\end{equation}
where the exponent $\mu\geq 0$ may vary based on the travel purpose. In general, it can be noticed that the higher $\mu$ the wider the angular span of the visual field, see Fig. \ref{fig.5}. Because of the lack of experimental data, the linear trend ($\mu=1$) is specifically retained here.

The perception point and the corresponding perceived density are evaluated as
\begin{equation}
	\x_p=\Psi[\rho]=\frac{1}{m_{R_s}}\lint_{R_s}\xibo\rho(\xibo,\,t)G(\alpha(\xibo-\x))\,d\xibo, \qquad
		\rho_p(\x,\,t)=\frac{m_{R_s}}{\vert R_s\vert},
\label{eq.xp.rhop.s4}
\end{equation}
where $m_{R_s}$, $\vert R_s\vert$ are the weighted mass of the crowd in $R_s$ and the weighted area of $R_s$, respectively:
\begin{equation}
	m_{R_s}=\lint_{R_s}\rho(\xibo,\,t)G(\alpha(\xibo-\x))\,d\xibo, \qquad
		\vert R_s\vert=\lint_{R_s}G(\alpha(\xibo-\x))\,d\xibo.
\label{eq.mRs.Rs.s4}
\end{equation}
\end{subequations}
In Eqs. \eqref{eq.xp.rhop.s4}, \eqref{eq.mRs.Rs.s4}, $\alpha=\alpha(\xibo-\x)$ is the angle that the vector $\xibo-\x$ forms with the desired direction of movement $\e_d(\x)$. The 1D counterparts of these formulas are obtained by setting $G\equiv 1$, because formally $\bar{\alpha}=0$. Then $x_p$ is the actual centre of mass of the crowd distributed in $[x,\,x+\delta]$, and $\rho_p(x,\,t)$ the actual average of $\rho$ therein.

The strategy just outlined relates to curious pedestrians: ensemble evaluation and synthesis are performed to both localise and evaluate the perceived density (e.g., in leisure or shopping activity).
\end{itemize}

\begin{figure}[!t]
\centering
\includegraphics[scale=0.87,clip]{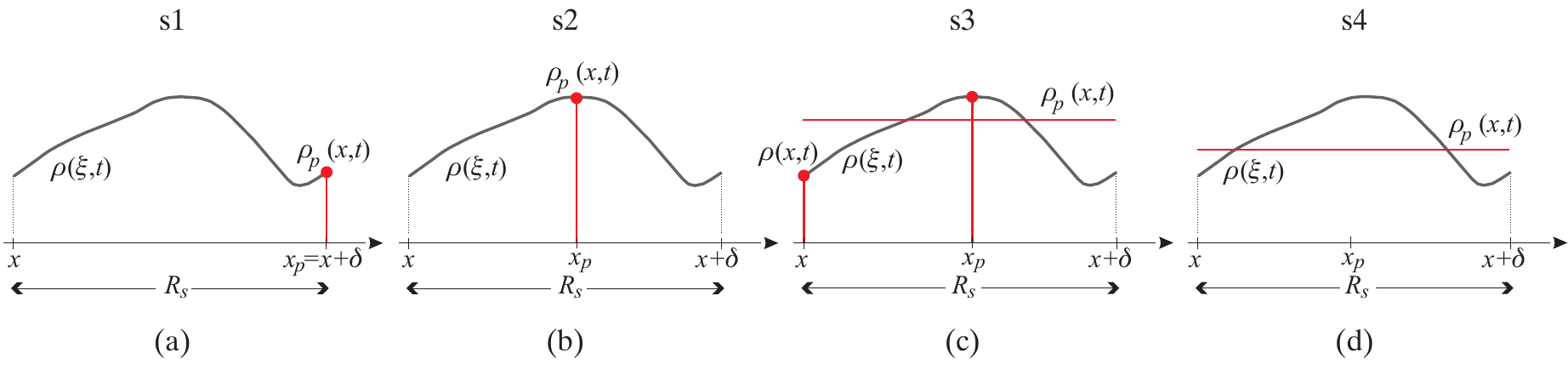}
\caption{1D examples of localisation strategies}
\label{fig.6}
\end{figure}

Pictorial representations of the above localisation strategies are provided in Fig. \ref{fig.6}, in a 1D domain and for a given spatial distribution of the actual crowd density in the sensory region.

\subsection{Qualitative analysis}
\label{two-3}
The qualitative analysis of the presented model develops in two steps. First, the diffusive effect induced by the non-local features of the model is analytically demonstrated by using a model equation in the 1D spatial domain. Second, the results obtained by using the four localisation strategies are numerically evaluated and discussed for a given true density distribution in both the 1D and the 2D case.

\subsubsection{Analysis of the diffusive effect induced by the non-local equations}
In order to study the effects introduced by the non-local in space fundamental relation, a model equation in the 1D spatial domain is analysed. Let us consider the mathematical model given by the 1D mass conservation equation, closed by a generic non-local fundamental relation:
\begin{equation}
\begin{cases}
	\dfrac{\partial\rho}{\partial t}+\dfrac{\partial\left(\rho v\right)}{\partial x}=0 \\[0.4cm]
	v=v(\rho_p).
\end{cases}
\label{eq.model_eq}
\end{equation}
Let us assume for simplicity that the space dislocation is evaluated by means of the simplest strategy among the presented ones (i.e., s1), so that $\rho_p(x,\,t)=\rho(x+\delta,\,t)$. Assuming that the extension of the sensory region is small compared to the characteristic spatial length (i.e., the domain size $L$), a first order Taylor expansion of $\rho$ gives:
\begin{equation}
\rho(x+\delta,\,t)\approx\rho(x,\,t)+\frac{\partial\rho}{\partial x}(x,\,t)\delta,
\label{eq.tay}
\end{equation}
so that
\begin{equation}
v(\rho_p)\approx v(\rho)+v'(\rho)\frac{\partial\rho}{\partial x}\delta,
\label{eq.v_approx}
\end{equation}
where $'$ will stand henceforth for derivative with respect to $\rho$. With the approximation \eqref{eq.v_approx}, and introducing the non-negative function $\nu(\rho)=-\rho v'(\rho)\delta$ (notice that $v$ is a decreasing function of the density, Eq. \eqref{eq.kla}), the mass conservation equation can be expressed in local form:
\begin{equation}
	\frac{\partial\rho}{\partial t}+\frac{\partial}{\partial x}(\rho v(\rho))
		-\frac{\partial}{\partial x}\left(\nu(\rho)\frac{\partial\rho}{\partial x}\right)=0.
\label{eq.G}
\end{equation}
It is worth pointing out that the function $\nu$ [m$^2$s$^{-1}$] plays the role of an equivalent kinematic viscosity induced by the non-locality parameter $\delta$. Multiplying Eq. \eqref{eq.G} by $\rho$ and then integrating on the spatial domain $\Omega=[0,\,L]$ gives
\begin{equation}
	\lint_0^L\frac{\partial\rho}{\partial t}\rho\,dx+
		\lint_0^L\frac{\partial}{\partial x}(\rho v(\rho))\rho\,dx-
			\lint_0^L\frac{\partial}{\partial x}\left(\nu(\rho)\frac{\partial\rho}{\partial x}\right)\rho\,dx=0,
\label{eq.iform.1}
\end{equation}
whence, setting $q(\rho)=\rho v(\rho)$, a further manipulation yields
\begin{equation}
	\frac{1}{2}\frac{d}{dt}\lint_0^L\rho^2\,dx+
		\lint_0^L q'(\rho)\frac{\partial\rho}{\partial x}\rho\,dx-
			\lint_0^L\frac{\partial}{\partial x}\left(\nu(\rho)\frac{\partial\rho}{\partial x}\right)\rho\,dx=0.
\label{eq.iform.2}
\end{equation}
Integrating by parts the second and the third term at the left-hand side, with periodic boundary conditions applied to $\rho$, Eq. \eqref{eq.iform.2} finally implies
\begin{equation}
	\frac{1}{2}\frac{d}{dt}\lint_0^L\rho^2\,dx=
		-\lint_0^L\nu(\rho)\left(\frac{\partial\rho}{\partial x}\right)^2\,dx\leq 0,
\label{eq.iform.3}
\end{equation}
hence the spatial $L^2$ norm of the density (i.e., its ``energy'') decreases in time as it typically happens in diffusion phenomena. In other words, the spatially non-local model does not only allow the pedestrian intelligent behaviour to be described but gives also the mass conservation equation formally a parabolic trend, which for instance prevents the formation of unrealistic shock waves.

Analogous conclusions are expected to hold qualitatively also for the other localisation strategies and in the 2D case.

\subsubsection{Effects of the different localisation strategies on the perceived density}
\begin{figure}[!t]
\centering
\includegraphics[width=0.8\textwidth,clip]{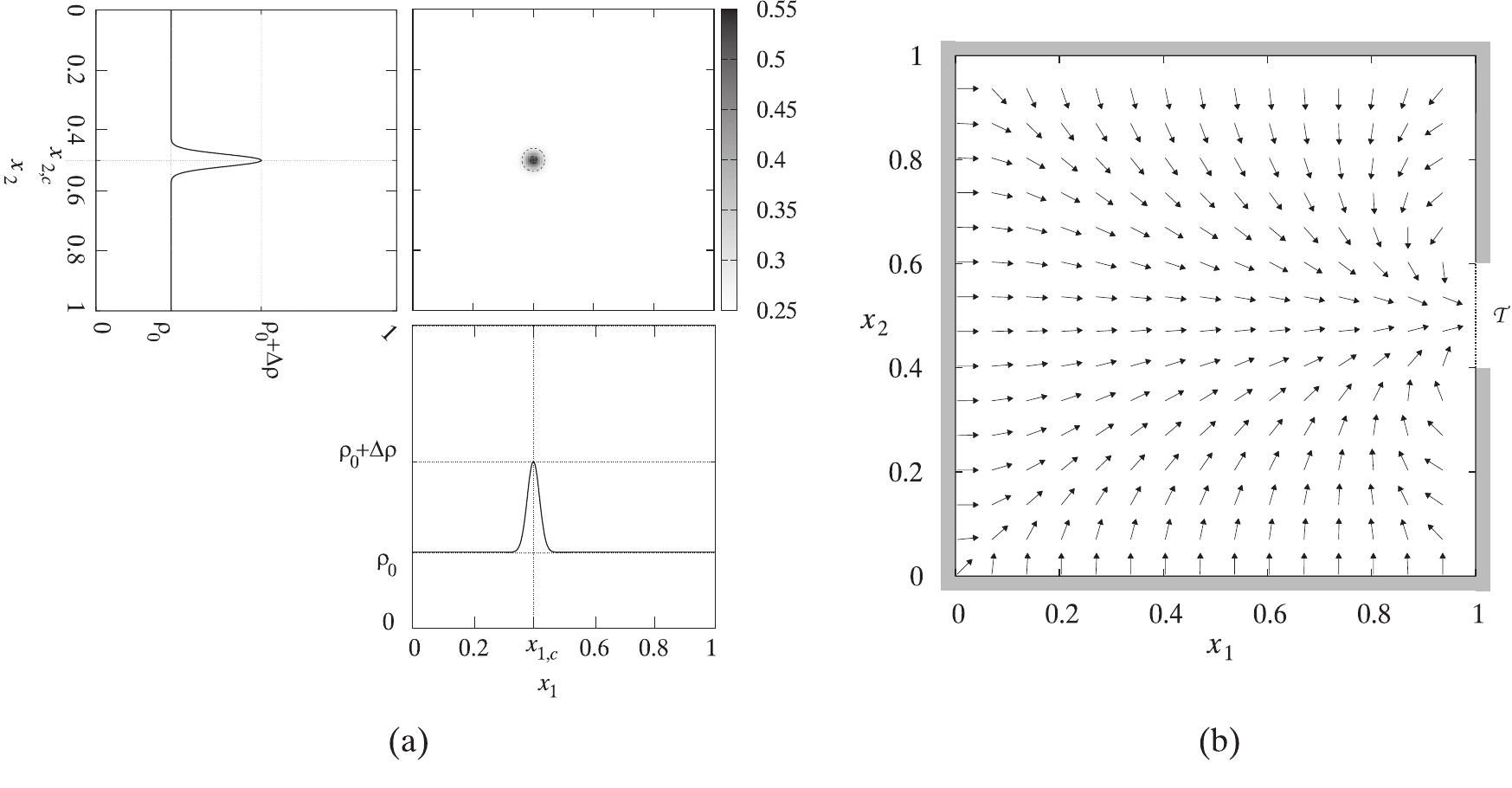}
\caption{(a) Filled contours and cross-sections of the instantaneous true density. (b) Desired direction $\e_d$ for the same test case}
\label{fig.7}
\end{figure}

The effects of the different localisation strategies described in Sect. \ref{two-2} are now evaluated in both the 1D and the 2D setting by applying them to a test case with a given density distribution. In the following, the non-dimensional form of the model is considered, meaning that length, density, and speed are rescaled with respect to the reference values $L$, $\rho_M$, and $v_M$. The instantaneous crowd density in the non-dimensional 2D domain $\Omega=[0,\,1]\times[0,\,1]$ is taken to be
\begin{equation}
	\rho(x_1,\,x_2)=\rho_0+\Delta{\rho}\exp\left[-\frac{(x_1-x_{1,c})^2+(x_2-x_{2,c})^2}{\ell^2}\right],
	\label{eq.gauss.2D}
\end{equation}
see Fig. \ref{fig.7}a: the reference constant value $\rho_0\geq 0$ is perturbed about the point $\x_c=(x_{1,c},\,x_{2,c})$ by a bell-shaped curve attaining the maximum value $\Delta{\rho}>0$, so that the density $\rho$ ranges on the whole from $\rho_0$ to $\rho_0+\Delta{\rho}$. The parameter $\ell>0$ defines the width of the bell. For the distribution in the 1D case over the non-dimensional interval $\Omega=[0,\,1]$ the cross-section at $x_2=x_{2,c}$ is used (continuous line in Fig. \ref{fig.9}). The coefficients in Eq. \eqref{eq.gauss.2D} are set to $\rho_0=0.25$, $\Delta{\rho}=0.3$, $\ell=1/35$, $x_{1,c}=0.4$, $x_{2,c}=0.5$. Finally, in the simulations the fundamental relation \eqref{eq.kla} is set for Europe and rush-hour traffic, that is, $\rho_M=6$ ped/m$^2$, $v_M=1.69$ m/s, and $\gamma=0.273\rho_M$. It is worth noticing that, given the previous value of $\ell$, the density bump is quite narrow, which reflects a strong non-equilibrium in the density distribution. In this situation, the ratio $\delta/\ell$ should be generally much larger than $1$, meaning that the sensory region might fully encompass the bump. Significant differences among the various perception strategies are therefore expected.

\begin{figure}[!t]
\centering
\includegraphics[width=\textwidth]{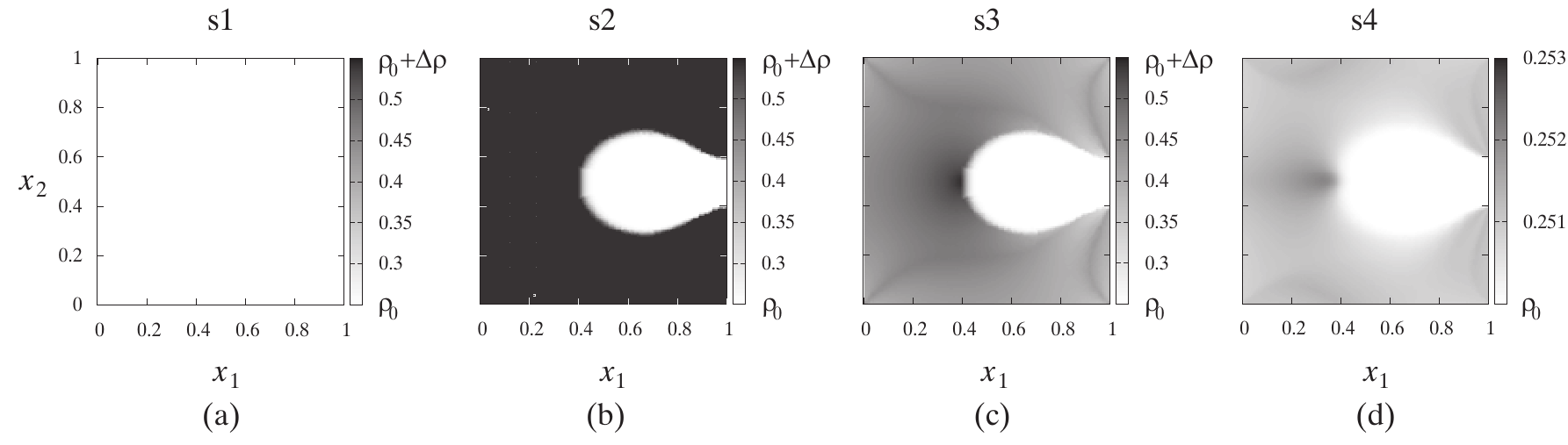} \\[0.3cm]
\caption{Filled contours of the perceived density $\rho_p$}
\label{fig.8}
\end{figure}

The perceived density $\rho_p$ resulting from each of the localisation strategies s1, \dots, s4 is shown in Fig. \ref{fig.8} for the 2D case and in Fig. \ref{fig.9} for the 1D case. Instead, in Fig. \ref{fig.10} the localisation mechanism underlying strategy s1 is pictorially represented in the 1D case as an example. The following considerations are in order.
\begin{enumerate}
\item[(i)] In strategy s1 the perceived density is constantly equal to $\rho_0$ throughout the domain in both the 1D and the 2D case (Fig. \ref{fig.8}a), which implies that pedestrians are not affected at all by the denser crowding about $\x_c$. The reason for this outcome is that the sensory length $\delta$ is everywhere quite large, because in a clear environment the maximum visual depth $\Delta_s$ is not obstructed by intermediate obstacles. As a consequence, the perception point located at the periphery of the sensory region (Eqs. \eqref{eq.xp.s1.2D}, \eqref{eq.xp.s1.1D}) never falls within the density bump if the latter is sufficiently narrow. This is particularly evident in the 1D setting: pedestrians located upstream the bump ($x<x_c$) tend to perceive the density condition downstream the bump (Fig. \ref{fig.10}).

\item[(ii)] In strategy s2 pedestrians are sensitive to the highest density they perceive within their sensory region $R_s$. In particular, as soon as the centre $\x_c$ of the bump falls in $R_s$ they react to the \emph{maximum} density in the domain, namely $\rho_0+\Delta{\rho}$. As a result, in the 2D setting a large area of $\Omega$ features a perceived density $\rho_p$ equal to $\rho_0+\Delta{\rho}$, possibly also in points far from $\x_c$ (Fig. \ref{fig.8}b) according to the local orientation of the desired direction of movement (Fig. \ref{fig.7}b). Conversely, in a smaller area immediately downstream the bump and in front of the target $\T$ the perceived density coincides with the true one, for the anisotropy of the sensory region makes pedestrians insensitive to what happens behind them. The analogy in the 1D setting is that the perceived density constantly equals $\rho_0+\Delta{\rho}$ for $x\leq x_c$, then it decreases to $\rho_0$ for $x>x_c$ following closely the profile of the true density.

\item[(iii)] In strategy s3 pedestrians are still affected by the highest density perceived within the sensory region, but in this case they are able to weight this information on the basis of the relative position of the perception point. The resulting 2D perceived density pattern (Fig. \ref{fig.8}c) is similar to that obtained with strategy s2, however in those points $\x$ whose sensory region encompasses $\x_c$ the maximum $\rho_0+\Delta{\rho}$ is attenuated, as it is averaged with $\rho_0$ according to the distance $r_p=\vert\x_c-\x\vert$. Specifically, using Eq. \eqref{eq.rhop.s3}, $\rho_p$ results of the order of $\rho_0+g(r_p)\Delta{\rho}$, which does not exceed $\rho_0+\Delta{\rho}$ because $g(r_p)\leq 1$. The 1D case highlights in particular that, for $x\leq x_c$, the perceived density gradually increases as the pedestrian position approaches the location of the peak of the true density. Conversely, for $x>x_c$ the trend is completely analogous to that of strategy s2.

\item[(iv)] In strategy s4 pedestrians operate a global synthesis of the mass of people distributed in the sensory region $R_s$, indeed their perceived density corresponds to an integral average of the true density within $R_s$. In the 2D case, this average also takes into account the lateral visual fading and the consequent relative enhancement of the frontal vision. The resulting perceived density pattern (Fig. \ref{fig.8}d, notice the change of colour scale with respect to the previous cases) is more homogeneous than that produced by strategies s2 and s3 (in the sense that the gap between the minimum and the maximum value of $\rho_p$ in $\Omega$ is much smaller than in the other two cases), nevertheless it is not constant like that returned by strategy s1 (that is, the bump of density in $\x_c$ is somehow felt). Notice however that the maximum of $\rho_p$ in $\Omega$ is definitely lower than that of $\rho$ as a consequence of the average process. The 1D case clarifies that pedestrians approaching the bump feel a much lower density than the peak value, whereas those located immediately downstream the bump react actually to a density lower than the local one, since they perceive the dispersion of the crowd in front.
\end{enumerate}

\begin{figure}[!t]
\centering
\includegraphics{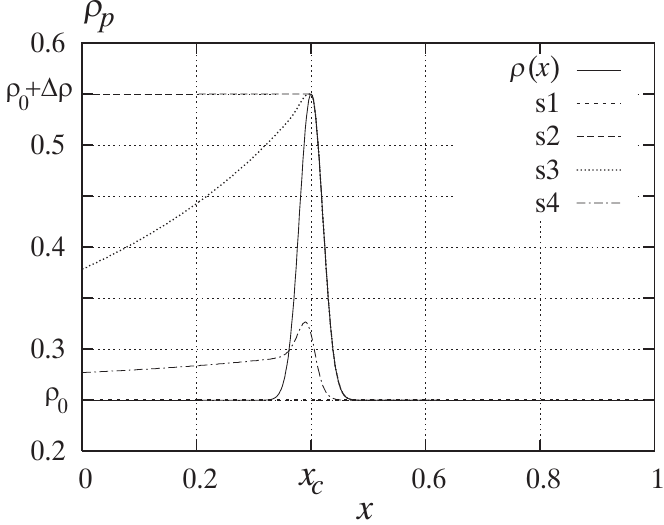} 
\caption{Perceived density according to the four strategies s1, \dots, s4}
\label{fig.9}
\end{figure}
\begin{figure}[!t]
\centering
\includegraphics{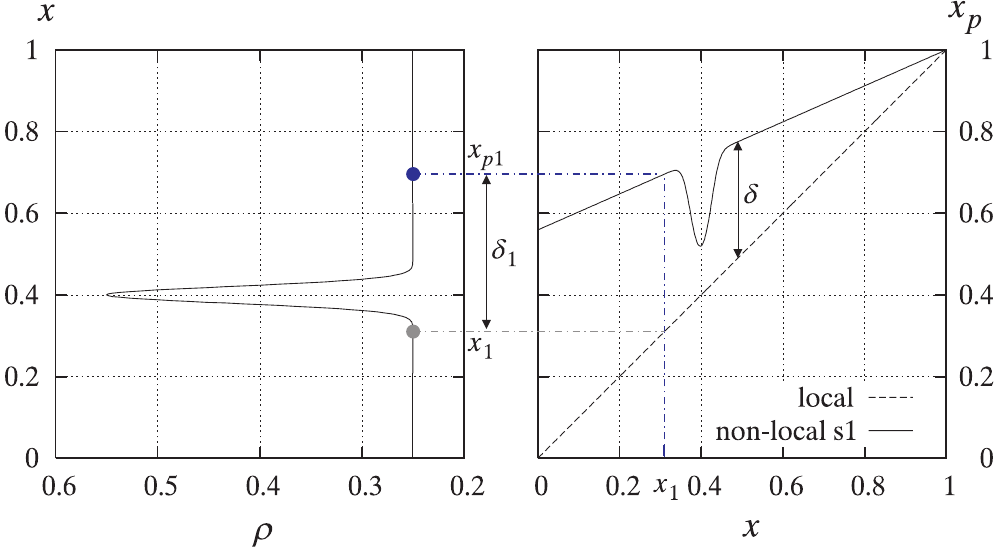} 
\caption{Example of localisation for strategy s1}
\label{fig.10}
\end{figure}

Comparing in particular the last three strategies, it can be noticed how the steep gradients of $\rho_p$ featured by s2 are gradually smoothed in s3 and dramatically smoothed in s4, as a consequence of less and less pointwise perception criteria. In general, all strategies seem plausible and suitable to address various applications implying different crowd traffic conditions.

\section{Real world applications}
\label{three}
In this section, the physical framework previously described is used to study the flow of a crowd in two specific real world situations: along a vibrating footbridge (1D domain) and in an underground station (2D domain) with obstacles and multiple exits. The most convenient modelling methods and numerical approaches are developed accordingly, in the frame of macroscopic first-order models. Numerical simulations are performed with the non-dimensional model: the crowd density and speed are rescaled with respect to $\rho_M$ and $v_M$, while lengths are rescaled with respect to the characteristic dimension of the walking area $L$. The velocity-density relation \eqref{eq.kla} is adapted for the case of Asia and rush-hour traffic: $\rho_M=7.7$ ped/m$^2$, $v_M=1.48$ m/s, and $\gamma=0.273\rho_M$.

\subsection{Case study 1: 1D model of crowd-structure interaction}
\label{three-one}
A 1D macroscopic model is suitable to describe specific kinds of crowd events in some walking facilities, such as dense crowds walking along walkway, corridors, and footbridges. In fact, the flow along such facilities is mainly one-dimensional because of their line-like geometry (i.e., the length is one or two orders of magnitude larger than the width of the walkway) and because pedestrians usually share the main objective of reaching a common target (for instance during particular events, such as opening days or demonstrations, and/or because of the specific function and location of the facility with respect to transport facilities).

\begin{figure}[!t]
\centering
\includegraphics[width=10cm]{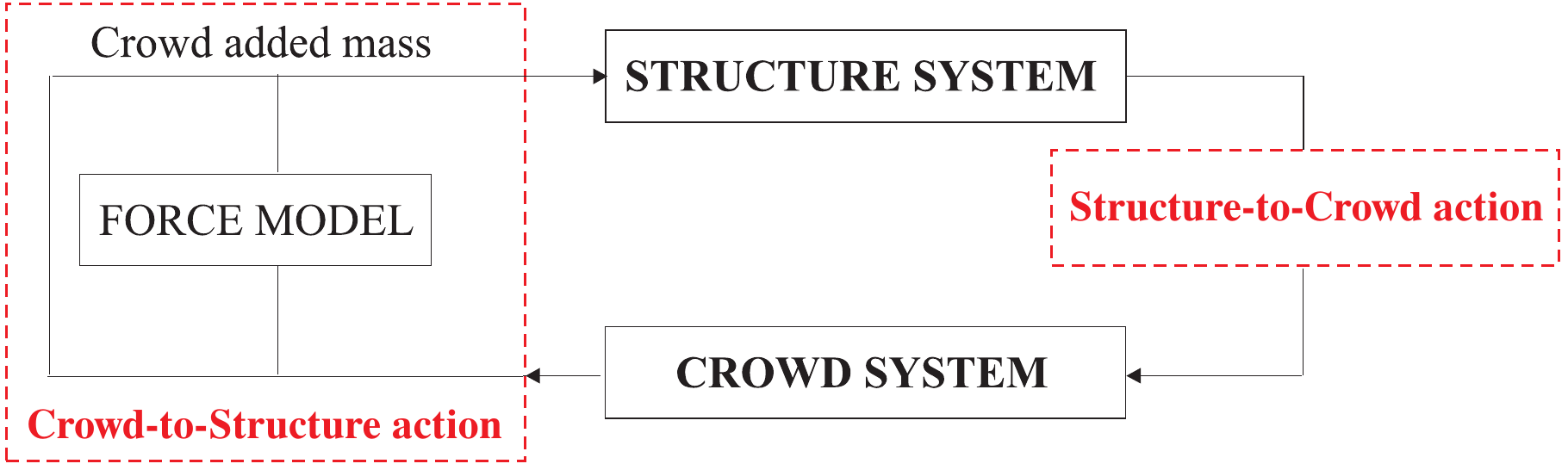}
\caption{Scheme of the crowd-structure coupled model} 
\label{fig.11}
\end{figure}

If the aim is to model the crowd flow along a lively structure, such as a slender footbridge, crowd-structure interaction should be properly modelled. In order to do so, the so-called partitioned approach \cite{park} in its differential form is applied to the coupled crowd-structure system, according to the framework schematised in Fig. \ref{fig.11}. The coupled model has been developed in several studies, summarised in \cite{venPLR,venJSV}: its detailed description is beyond the scope of the present paper, and only some fundative concepts are introduced in the following to allow the reader to interpret the presented results.

With respect to the physical framework previously introduced, some additional modelling is needed in order to describe the action exerted by the structure subsystem on the crowd subsystem. The following assumptions are introduced:
\begin{enumerate}
\item[(i)]the motion of the structure, expressed by the envelope of its lateral acceleration $\tilde{\ddot z}(x,\,t)$, reduces the walking speed;

\item[(ii)] pedestrians are not affected by the motion of the structure if the acceleration is under the threshold of motion perception $\ddot z_c$ ($\ddot z_c=0.1$ m/s$^2$ in the following);

\item[(iii)] pedestrians react to the motion of the structure non-locally in time, i.e., with a time delay $\hat{\tau}_{1}$ that is expected to be greater than the time interval between two succeeding steps (in the following $\hat{\tau}_{1}=1$ s);

\item[(iv)] after pedestrians have stopped at time $t_s$ because of excessive acceleration $\ddot z_M$ ($\ddot z_M=2.1$ m/s$^2$), a stop-and-go time interval $\hat{\tau}_{2}$
($\hat{\tau}_{2}=5$ s) should elapse before they start walking again.
\end{enumerate}
Notice that $\hat{\tau}_{1}$ and $\hat{\tau}_{2}$ can be ascribed to the reflex and volitional time delays mentioned in Sect. \ref{two}, respectively. However, in the present context they are referred to pedestrian perception of the platform motion by means of their feet and legs, rather than by means of the visual perception.

The mathematical model describing the 1D crowd subsystem is given by the mass conservation equation, closed by a constitutive law linking the walking speed $v$ to the perceived crowd density $\rho_p$:
\begin{equation}
\begin{cases}
	\dfrac{\partial\rho}{\partial t}+\dfrac{\partial(\rho v)}{\partial x} = 0 \\[0.4cm]
	v=v(\rho_p)g(\tilde{\ddot z}),
\end{cases}
\label{eq.frame_pde}
\end{equation}
where the function $v(\rho_p)$ is as in Eq. \eqref{eq.kla} and $g(\tilde{\ddot{z}})=g(\tilde{\ddot{z}}(x,\,t-\hat{\tau}_{1}))\in[0,1]$ is a corrective function with the following qualitative trend:
\begin{equation}
	g(\tilde{\ddot{z}})=
	\left\{
	\begin{array}{ll}
		1 & \text{if\ } \tilde{\ddot{z}}\leq\ddot{z}_c\wedge
			(t\leq t_s\vee t\geq t_s+\hat{\tau}_{2}) \\[0.2cm]
		\dfrac{\ddot{z}_M-\tilde{\ddot{z}}(x,\,t-\hat{\tau}_{1})}{\ddot{z}_M-\ddot{z}_c} &
			\text{if\ } \ddot{z}_c<\tilde{\ddot{z}}<\ddot{z}_M\wedge
				(t\leq t_s\,\vee\,t\geq t_s+\hat{\tau}_{2}) \\[0.4cm]
	  	0 & \text{if\ } \tilde{\ddot{z}}\geq\ddot{z}_M\wedge
	  		(t\leq t_s\vee t_s<t<t_s+\hat{\tau}_{2}).
	  \end{array}
	  \right.
\label{eq.vz}
\end{equation} 
Model \eqref{eq.frame_pde} must be supplemented by boundary conditions at $x=0$ and $x=L$.

The crowd-to-structure action takes place in two ways. First, the crowd-structure total mass $m$ is constantly updated by adding the pedestrian mass $m_c$ to the structural mass $m_s$. Second, the lateral force $F(x,\,t)$ exerted by pedestrians is expressed as a function of both $\rho$ and $\tilde{\ddot{z}}$. The interested reader can refer to \cite{venJSV} for a  description of the macroscopic force model.

Finally, the structure subsystem is modelled as a non-linear three dimensional (3D) damped dynamical system, whose equation of motion can be written as:
\begin{equation}
\left[ m_s + m_c(\rho) \right] \frac{\partial^2 s}{\partial t^2}+{\mathscr C}\left[ \frac{\partial s}{\partial t}\right]+ {\mathscr L}\left[ s\right] = F(\rho,\tilde{\ddot{z}}),
\label{eq.moto}
\end{equation}
where $s=s(\x,t)$ is the structural displacement; $\mathscr {C}$ and $\mathscr L$ are the damping and stiffness operators, respectively; $F(\rho,\tilde{\ddot{z}})$ is the applied lateral force introduced above.

\subsubsection{Numerical approach}
In order to solve the crowd model described by the scalar non-linear Eq. \eqref{eq.frame_pde}, suitable numerical schemes for time and space discretisation are required. The classical explicit Euler scheme is adopted for the advancement in time, while the spatial derivative is discretised using a finite-volume approximation with first-order monotonic upwind scheme, according to Godunov's theorem \cite{god}. Therefore, the discretised mass conservation equation at the $j$-th cell becomes:
\begin{equation}
	\frac{\rho_j^{n+1}-\rho_j^{n}}{\Delta t}=-\dfrac{q_{j+1/2}-q_{j-1/2}}{\Delta x}
	\label{eq.TD}
\end{equation}
where $\Delta x$ and $\Delta t$ are the uniform grid size and the time step, respectively, and $q_{j\pm 1/2}$ are the numerical fluxes at the cell interfaces.

In spite of the classical numerical approach adopted, some attention must be paid in dealing with the non-local features of the proposed model. In particular, the above-mentioned Godunov's first-order upwind scheme \cite{lev} requires two main adjustments.

First, let us recall the well known Courant-Friedrichs-Lewy (CFL) stability condition:
\begin{equation}
	\sup\frac{\Delta t}{\Delta x}\vert v_c\vert \leq 1,
\label{eq.CFL}
\end{equation}
which involves the convection velocity $v_c$. Under the assumption that, for all fixed $t\geq 0$, the function $\rho$ is a diffeomorphism, the expression of $v_c$ in the non-local model is
\begin{equation}
	v_c(\rho,\,\rho_p)=v(\rho_p)+\rho v'(\rho_p)\frac{\partial\rho_p}{\partial x}\frac{\partial x}{\partial \rho}.
	\label{v_conv_disl}
\end{equation}
In order to define the fluxes $q_{j\pm 1/2}$ appearing in Eq. \eqref{eq.TD}, the convection velocity must be numerically evaluated at the cell interfaces. Adopting a linear interpolation scheme to compute the local density $\rho_{j+1/2}$ and the non-local density $\rho_{p,\,j+1/2}$ from their respective values in the cells $j$ and $j+1$, the discrete convection velocity at the cell interface $j+1/2$ reads
\begin{equation}
	v_{c,\,j+1/2}=v(\rho_{p,\,j+1/2})+\rho_{j+1/2}v'(\rho_{p,\,j+1/2})\frac{\Delta\rho_p}{\Delta\rho},
	\label{v_conv_disl_disc}
\end{equation}
where $\Delta\rho_p=\rho_{p,\,j+1/2}-\rho_{p,\,j}$ and $\Delta\rho=\rho_{j+1/2}-\rho_j$. Two important features of Eq. \eqref{v_conv_disl_disc} are worth a mention:
\begin{itemize}
\item the sign of the ratio $\Delta\rho_p/\Delta\rho$ affects the sign of the convection velocity and the selection of the upwind node in turn. In other words, the sign of $v_c$ cannot be determined by simply referring to the so-called capacity density value in the speed-density constitutive law illustrated in Fig. \ref{fig.2}a;

\item the value of the ratio $\Delta\rho_p/\Delta\rho$ affects the CFL condition. In particular, $v_c$ may tend to infinity for vanishing $\Delta\rho$, which would make it impossible to select a discrete time step $\Delta{t}$ fulfilling the CFL condition \eqref{eq.CFL}.
\end{itemize}
A threshold $\epsilon>0$ is set on $\left\vert\Delta\rho\right\vert$: if $\left\vert\Delta\rho\right\vert>\epsilon$ then Eq. \eqref{v_conv_disl_disc} holds, otherwise $\left\vert\Delta\rho\right\vert=\epsilon$ is imposed, because pedestrians are supposed to be unable to evaluate differences in local density smaller than $\epsilon$. In the following $\epsilon=10^{-4}$ is chosen and, in order to save computational resources, an adaptive time step is adopted.

Second, the numerical flux at the interface requires a non-classical definition in order to pursue entropic weak solutions \cite{lev}:
\begin{equation}
	q_{j+1/2}=
	\begin{cases}
		\displaystyle{\min_{\eta\in[0,\,1]}}q(\rho^\eta,\,\rho_p^\eta) &
			\text{if\ } \rho_j\leq\rho_{j+1} \\
		\displaystyle{\max_{\eta\in[0,\,1]}}q(\rho^\eta,\,\rho_p^\eta) &
			\text{if\ } \rho_j>\rho_{j+1},
	\end{cases}
\label{GFF}
\end{equation}
where $\rho^\eta=(1-\eta)\rho_j+\eta\rho_{j+1}$ and $\rho_p^\eta=(1-\eta)\rho_{p,\,j}+\eta\rho_{p,\,j+1}$ are linear interpolations of the values of $\rho$, $\rho_p$ at the sides of the interface $j+1/2$.

The Finite Element Method (FEM) is employed for the space discretisation of the 3D structural multi-degree-of-freedom model, while its advancement in time is obtained by means of the classic Newmark method.

The two subsystems are characterised by non-matching grids in space thanks to the differential partitioning adopted, while they share the same discretisation in time \cite{park}.

\subsubsection{Results}
The proposed model is applied to the simulation of the crowd dynamics along a real world cable-stayed footbridge located at Toda City, Japan. The footbridge connects a boat race stadium to a bus terminal: at the end of the races, the bridge is crossed by the spectators who leave the stadium to reach the terminal. During these events, the crowd density attains 1.5 ped/m$^2$ \cite{fuji}: in these circumstances, the bridge deck experiences excessive lateral vibrations due to crowd-structure interaction.

The crowd flow is simulated under three different conditions (referred to as ``set-up'' in the following) of increasing complexity: (i) along the motionless deck (i.e., only crowd dynamics is studied); (ii) with an imposed deck lateral motion (i.e., only the structure-to-crowd action is accounted for); (iii) in the complete two-way crowd-structure interaction. These simulations are aimed at discussing the influence of the non-local pedestrian behaviour on the crowd dynamics, the effects of the lateral motion of the platform on the crowd flow, and the effects of the non-local pedestrian behaviour on the structural response of the footbridge. Only one localisation strategy, namely s3, is adopted in this application.

\begin{figure}[!t]
\centering
\includegraphics[width=\textwidth,clip]{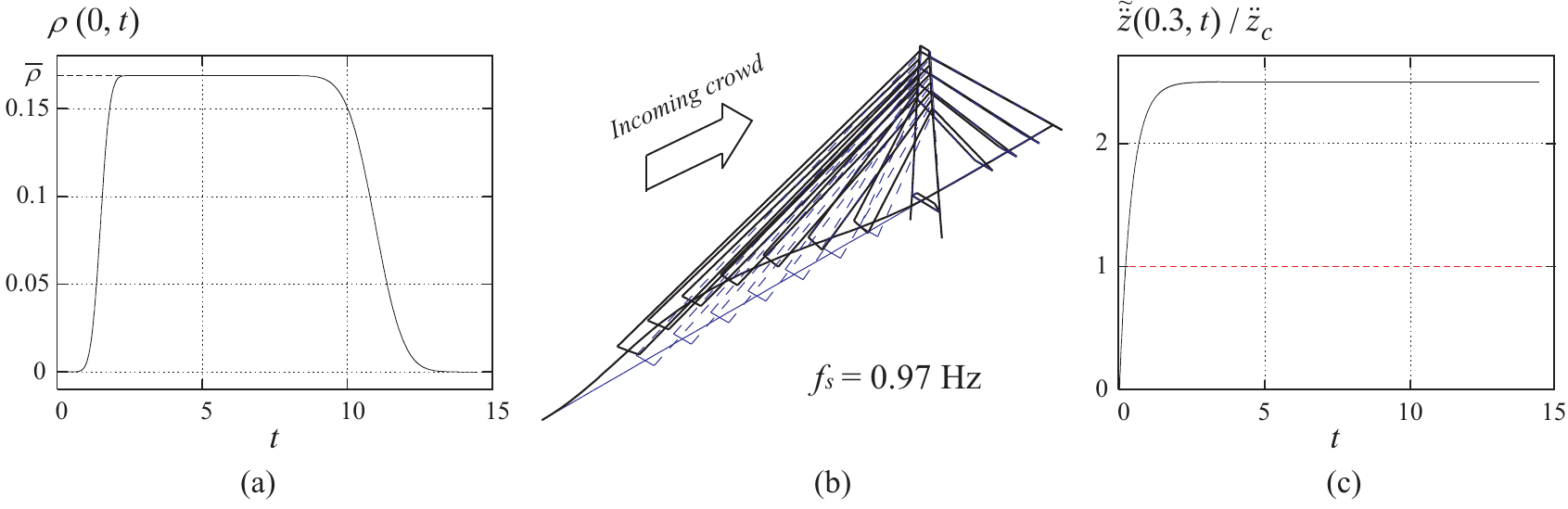}
\caption{T-bridge: (a) inlet b.c. on the crowd density, (b) structural FEM model, first lateral mode shape and related frequency $f_s$, (c) imposed envelope of lateral acceleration of the deck in $x=0.3$}
\label{fig.12}
\end{figure}

The selected inlet boundary condition (b.c.) for the crowd density (Fig. \ref{fig.12}a) exhibits a steady-state regime, with $\bar{\rho}=0.17$ (corresponding to 1.3 ped/m$^2$), comprised between two transient regimes marking the beginning and the end of the stadium evacuation. It should be noticed that the density decreases in a smoother way than the initial increase. This is due to the assumption that the stadium evacuation abruptly starts at the end of the boat race with a sudden increase in the crowd density, while the complete evacuation is expected to be smoother in time.

The footbridge structure is characterised by a two-span continuous steel box girder, a two-plane multistay cable system, and a reinforced concrete tower. The total bridge length $L$ is about 180 m and the road deck width is 5.25 m. The deck mass per square metre is 800 kg/m$^2$ and the damping ratio is approximately 0.7\%. The tower and the deck are modelled with elastic beam elements, while each cable is modelled using a single truss element (Fig. \ref{fig.12}b). The damping is modelled via Rayleigh stiffness proportional damping. More details about the structural model can be found in \cite{venJSV}.

The envelope $\tilde{\ddot z}(x,\,t)$ of the lateral acceleration of the deck within set-up (ii) is obtained as $\bar{\ddot z}(1-\exp(-\beta t))\phi_1(x)$, where $\bar{\ddot z}=0.25$ m/s$^2$ is the maximum value of the acceleration, $\beta=0.02$ and $\phi_1(x)$ is the first lateral mode shape in Fig. \ref{fig.12}b. The time history in $x=0.3$ is represented in Fig. \ref{fig.12}c.

\begin{figure}[!t]
\centering
\includegraphics[width=\textwidth]{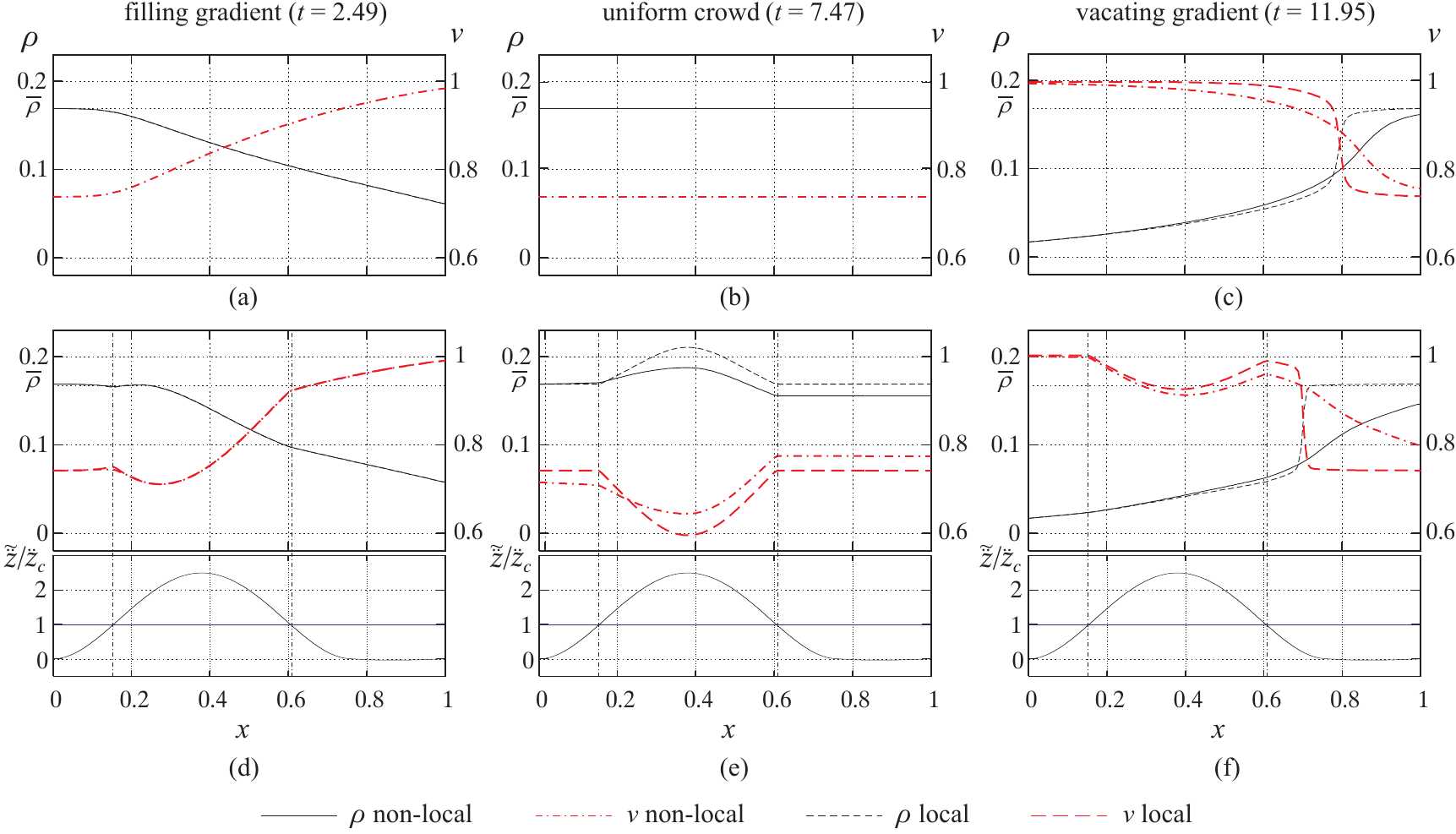}
\caption{Crowd dynamics: instantaneous density and speed for motionless (a-c) and moving (d-f) platform}
\label{fig.13}
\end{figure}

Figure \ref{fig.13} summarises some results referred to the set-ups (i) (Figs. \ref{fig.13}a-c) and (ii) (Figs. \ref{fig.13}d-f) in terms of the instantaneous crowd density and speed at three time instants corresponding to different regimes: the transient one, in which the crowd is gradually filling the deck span (called ``filling gradient''), the ``uniform crowd'' regime, in which an equilibrium condition is expected along the motionless deck, and the final transient regime, in which the crowd density gradually decreases at the footbridge entrance (``vacating gradient'' in the following). For the set-up (ii), the instantaneous lateral acceleration of the deck is also plotted (Figs. \ref{fig.13}d-f). Notice that in the interval $[\underline{x},\overline{x}]$ the acceleration exceeds the threshold value $\ddot{z}_c$.

\begin{figure}[!t]
\centering
\includegraphics{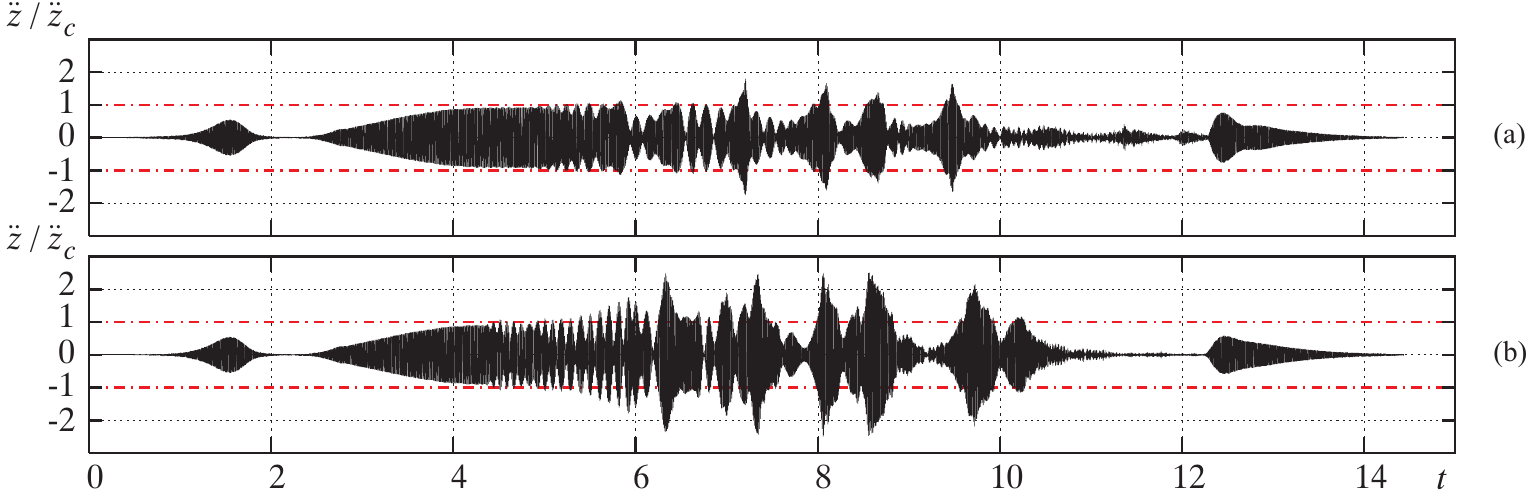} 
\caption{Crowd-structure interaction: deck lateral acceleration at $x=0.3$ with local (a) and non-local (b) model}
\label{fig.14}
\end{figure}

The following considerations can be synthetically made.
\begin{itemize}
\item In the filling gradient regime, no effects of the non-local model arise on either $\rho$ or $v$. Indeed, the adopted strategy s3 implies that the (discrete) perceived density equals the local one whenever the latter is monotonically decreasing. In the set-up (ii) (Fig. \ref{fig.13}d), the effects of the motion of the platform on the walking speed are clearly visible in the span interval in which the lateral acceleration is higher than $\ddot{z}_c$: the higher the latter, the lower the walking speed.

\item In the uniform crowd regime, local and non-local models provide the same results in set-up (i) (Fig. \ref{fig.13}b), as expected in equilibrium conditions. The combined effects of platform motion and non-local perception are highlighted in set-up (ii) (Fig. \ref{fig.13}e). It is worth recalling that the walking speed is affected both by the local platform motion and, in the non-local model, by the perceived density.
As a consequence, for $x<\underline{x}$ the two models predict different walking speed values in correspondence of the same density value. Indeed, in the non-local model the walking speed depends on the density perceived in a sensory region which includes part of the interval $[\underline{x},\overline{x}]$, where the platform acceleration exceeds $\ddot{z}_c$. For analogous reasons, along the interval $[\underline{x},\overline{x}]$ the platform motion and the anisotropic (forward) perception justify the scatter in the maximum (minimum, resp.) value of $\rho$ ($v$, resp.) and their skewed spatial distribution in the non-local model. In particular, the crossing of the two speed curves at $x \approx 0.23$ implies that there the local density is equal to the perceived one.

\item In the vacating gradient regime, the non-local model avoids the shock wave in the density distribution, coherently with the results of the qualitative analysis (Sect. \ref{two-3}). In addition, the platform motion (set-up (ii), Fig. \ref{fig.13}f) reduces the magnitude of the convection velocity in both models, so that the vacating wave front is shifted backwards with respect to set-up (i) (Fig. \ref{fig.13}c).
\end{itemize}

Finally, the structural responses predicted by the local and non-local crowd model in set-up (iii) are compared in Fig. \ref{fig.14} in terms of time histories of the lateral acceleration at the point $x=0.3$. On the one hand, the structural responses predicted by the two models share the same general trend in time: in particular, the peak lateral acceleration is reached during the uniform crowd regime. On the other hand, the acceleration peak value predicted by the non-local model is approximatively 1.4 times greater than the one predicted by the local model. A detailed investigation of the reasons for this difference would require a full analysis of the adopted force model, which is beyond the scope of the present paper. An example of the relationships among the structural response, the pedestrian force, and the crowd density and velocity can be found in \cite{venJSV}. In the present context, it is worth stressing that this difference is relevant in the engineering practice of footbridge design, because the lateral acceleration affects both pedestrian comfort and footbridge serviceability \cite{ziv}.

\subsection{Case study 2: 2D model of pedestrian flow in built environments}
\label{three-two}
To approach the modelling of 2D pedestrian flows, a technique recently introduced in \cite{piccoli2009pfb,piccoli2008tem}, which slightly departs from the direct application of hyperbolic equations, is adopted. In fact, 2D hyperbolic equations would involve analytical and numerical difficulties, such as the correct definition of the convection velocity in presence of non-local flux and/or in domains with obstacles, that are instead bypassed by using the formalism introduced in the above cited papers.

The starting point is the mass conservation equation \eqref{Lf-mc}, which in Lagrangian form reads:
\begin{equation}
\frac{d}{dt}\lint_{\gamma(\E,\,t)}\rho(\xibo,\,t)\,d\xibo=0,
\label{eq.2D-rho}
\end{equation}
where $\gamma(\E,\,t)$ is the configuration at time $t>0$ of the material sub-domain $\E\subseteq\Omega$. In more detail, the \emph{flow map} $\gamma$ is such that $\x=\gamma(\X,\,t)$ is the position occupied at time $t>0$ by the point initially located in $\X$, therefore Eq. \eqref{eq.2D-rho} expresses the conservation of the crowd mass contained in $\E$ along the spatio-temporal patterns of the system. The flow map is related to the velocity $\vv=\vv(\x,\,t)$ by
\begin{equation}
	\begin{cases}
		\dfrac{\partial\gamma}{\partial t}(\X,\,t)=\vv(\gamma(\X,\,t),\,t) \\
		\gamma(\X,\,0)=\X.
	\end{cases}
	\label{eq.2D-Cauchy_gamma}
\end{equation}

A major difference with respect to the 1D case is that in the present 2D setting it is possible, indeed almost essential, to take into account obstacles scattered in the walking area. These make some portions of the walking surface inaccessible to pedestrians, hence from the modelling point of view they can be understood as internal holes to the domain $\Omega$. The desired direction $\e_d$ must then be modelled coherently with these constraints, so as to point towards the target $\T$ while bypassing intermediate holes. The assumption that pedestrians \emph{a priori} know the walking area (Sect. \ref{two-2}) implies that $\e_d$ depends on the geometry of the domain only. In other words, the desired direction is not affected by either the perception process or the dynamics of the interactions, hence it can be regarded to all purposes as a datum of the problem. A possibility, which allows to deal with quite complex domains in a relatively easy manner, is to obtain $\e_d$ as the (normalised) gradient of a scalar potential $u=u(\x):\Omega\to\R$ satisfying Laplace's equation:
\begin{align}
	& \Delta{u}=0 \quad \text{in\ } \Omega,
	\label{eq.laplace} \\[0.1cm]
	& \e_d(\x)=\frac{\nabla{u}(\x)}{\vert\nabla{u}(\x)\vert}.
	\label{eq.2D-ed}
\end{align}
Boundary conditions are required along $\partial\Omega$, including the internal boundaries delimiting the obstacles, in order to identify the attracting destination as well as the perimeter walls that pedestrians cannot cross. A detailed analysis of these issues can be found in \cite{piccoli2009pfb}.

\subsubsection{Numerical approach}
\label{num.appr.2D}
In order to solve numerically the equations of the model it is necessary to design a suitable scheme for the approximation in time and space of the density $\rho$. In \cite{piccoli2008tem} a discretisation method of the mass conservation equation in the form \eqref{eq.2D-rho} is proposed and proved to guarantee properties of conservativeness, positiveness, and spatial accuracy. Before outlining the main underlying ideas, it is worth pointing out that this method is not concerned with the solution of Eqs. \eqref{eq.laplace}, \eqref{eq.2D-ed}, which instead can be performed \emph{a priori} by whatever technique suitable for elliptic equations (e.g., Finite Element Method).

Let us introduce a time discretisation $t_n=n\Delta{t}$, $n\in\mathbb{N}$, for a certain (possibly adaptive) time step $\Delta{t}>0$, then let us describe the action of the flow map $\gamma$ during one time step ($n\to n+1$) by means of a \emph{one-step flow map} $\gamma_n:\Omega\to\Omega$ in the following way: if $\x$ is the position occupied at time $t_n$ by the point initially located in $\X$ (in other words, $\x=\gamma(\X,\,t_n)$), then
\begin{equation}
	\gamma_n(\x)=\gamma(\X,\,t_{n+1}).
\end{equation}
A natural approximation of the one-step flow map, coming from an explicit Euler discretisation in time of Eq. \eqref{eq.2D-Cauchy_gamma}, is
\begin{equation}
	\gamma_n(\x)\approx\x+\vv_n(\x)\Delta{t}
	\label{eq.gamma.n}
\end{equation}
for $\vv_n(\x)=\vv(\x,\,t_n)$, as if the velocity were constant during one time step. Likewise, an explicit Euler discretisation in time of Eq. \eqref{eq.2D-rho} produces
\begin{equation}
	\lint_E\rho_{n+1}(\xibo)\,d\xibo=\lint_{\gamma_n^{-1}(E)}\rho_n(\xibo)\,d\xibo,
	\label{eq.2D-pushfwd_rho}
\end{equation}
where $\rho_n(\x)$ denotes (an approximation of) the density $\rho(\x,\,t_n)$ and $E$ is any subset of the domain $\Omega$.

As far as the spatial discretisation is concerned, let us assume that:
\begin{itemize}
\item the domain $\Omega$ is a rectangle,

\item the domain is partitioned with an orthogonal uniform Cartesian grid $\{E_{ij}\}$ consisting of $N_1\times N_2$ elements in the horizontal and vertical direction, respectively, each $E_{ij}$ being a square of characteristic edge's length $\Delta{x}>0$,

\item each obstacle can be described as the union of several grid cells, so as to be encompassed by the partition.
\end{itemize}
The density $\rho_n$ is approximated by a piecewise constant function taking the value $\rho^n_{ij}\geq 0$ in the grid cell $E_{ij}$. An analogous discretisation is applied also to the velocity $\vv_n$, which in general is not known exactly for it depends on the unknown $\rho_n$ through the interaction component $\e_i$. The approximate value of $\vv_n$ in $E_{ij}$ is denoted $\vv_{ij}^n\in\R^2$. This induces a further discretisation on the one-step flow map \eqref{eq.gamma.n}, that ultimately becomes a piecewise translation on the numerical grid (because the velocity is now constant in each cell). Plugging the approximate densities into Eq. \eqref{eq.2D-pushfwd_rho} and choosing $E=E_{ij}$ finally yields
\begin{equation}
	\rho^{n+1}_{ij}=\frac{1}{h^2}\sum_{k=1}^{N_1}\sum_{l=1}^{N_2}\rho^n_{kl}\vert E_{ij}\cap\gamma_n(E_{kl})\vert,
	\label{eq.2D-num_scheme}
\end{equation}
where $\vert\cdot\vert$ denotes the area of its argument.

Equation \eqref{eq.2D-num_scheme} provides an explicit numerical scheme for determining the new coefficients $\{\rho_{ij}^{n+1}\}_{i,j}$ from the knowledge of the coefficients $\{\rho_{ij}^n\}_{i,j}$ at the previous time step.  The proposed spatial discretisation is accurate with respect to the level of refinement of the numerical grid provided the parameters $\Delta{x}$, $\Delta{t}$ are linked by the following CFL-like condition \cite{piccoli2008tem}:
\begin{equation}
	\max_{i,j}\frac{\Delta{t}}{\Delta{x}}\vert\vv_{ij}^n\vert\leq 1.
	\label{eq.CFL.2D}
\end{equation}
Condition \eqref{eq.CFL.2D} compares well with Eq. \eqref{eq.CFL}, with the remarkable difference that it involves the values of the flux velocity rather than those of the convection velocity, so that the latter need not be determined.

\begin{figure}[!t]
\begin{center}
\includegraphics[width=\textwidth,clip]{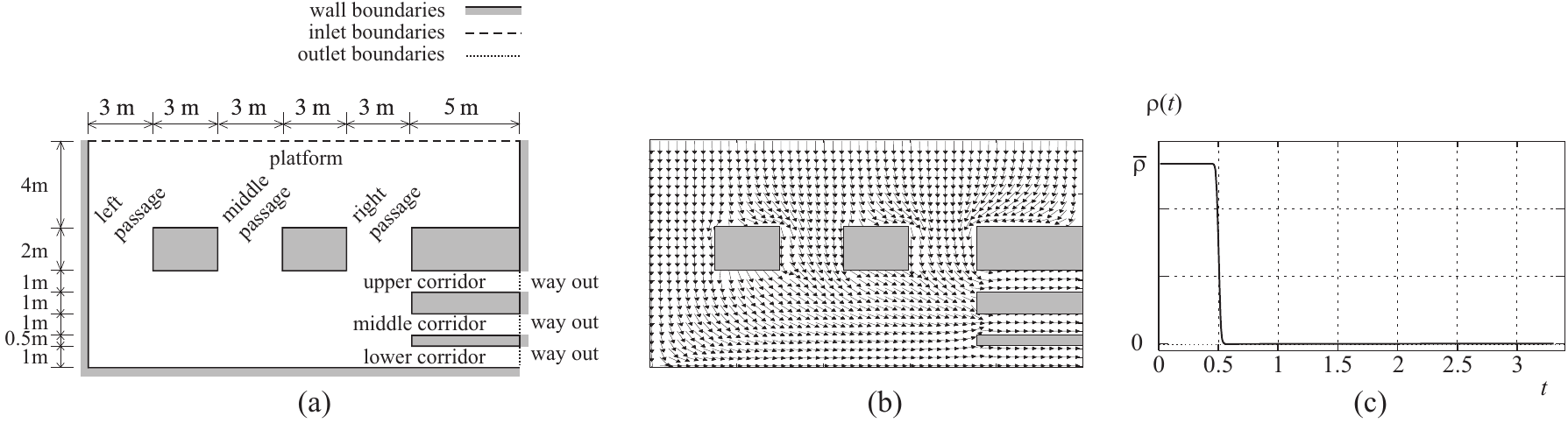}\\
\caption{(a) The map of the underground station used for numerical simulations. (b) The desired direction field $\e_d$. (c) Inlet b.c. on the crowd density}
\label{fig.15}
\end{center}
\end{figure}

\subsubsection{Results}
\label{sect.sim.2D}
In this section the 2D model is used to simulate a crowd leaving an underground station. The main goal is to investigate the effect of the various perception strategies in unsteady conditions and in built environments with obstacles. Figure \ref{fig.15}a shows the map of the underground station considered in the simulations, the grey blocks representing pillars and handrails delimiting the platform and some corridors, respectively. It is assumed that the left, right, and bottom edges of the walking area are perimeter walls, that pedestrians cannot cross but for the exits at the end of the corridors.

Although it does not reproduce any particular station, the map features some basic characteristics quite common to many real stations. Pedestrians getting off the trains access the station from the platform, then they aim at the exits placed at the end of three corridors. The corresponding desired direction field $\e_d$ is obtained accordingly by means of Eqs. \eqref{eq.2D-ed}, \eqref{eq.laplace}, see Fig. \ref{fig.15}b. Specifically, the potential $u$ is set to the maximum value ($u=1$) at the end of the corridors. Conversely, the normal derivative of $u$, namely the normal component of $\e_d$, is prescribed at the boundaries of the pillars and of the perimeter walls in such a way that pedestrians do not tend to cross the edges while possibly sliding along them. The direction of movement $\e_v$ is determined according to Eq. \eqref{eq.dir}, with the parameter $\theta$ set to $\theta=0.7$ to avoid reversed flow.

The unsteady inlet b.c. on the crowd density is plotted in Fig. \ref{fig.15}c, the maximum density value being $\bar{\rho}=0.02$ (corresponding to 0.15 ped/m$^2$): such a time history is intended to represent the arrival of a train at the station and its subsequent restart. It is worth noting that the low value of $\bar{\rho}$ is chosen in order to avoid the congested regime ($\bar{\rho}<\rho_{ca}$, Fig. \ref{fig.2}b) in the platform and along the passages, and to allow a sound comparison of the strategy effects on the free flow approaching the corridors.

\begin{figure}[!t]
\begin{center}
\includegraphics[width=\textwidth,clip]{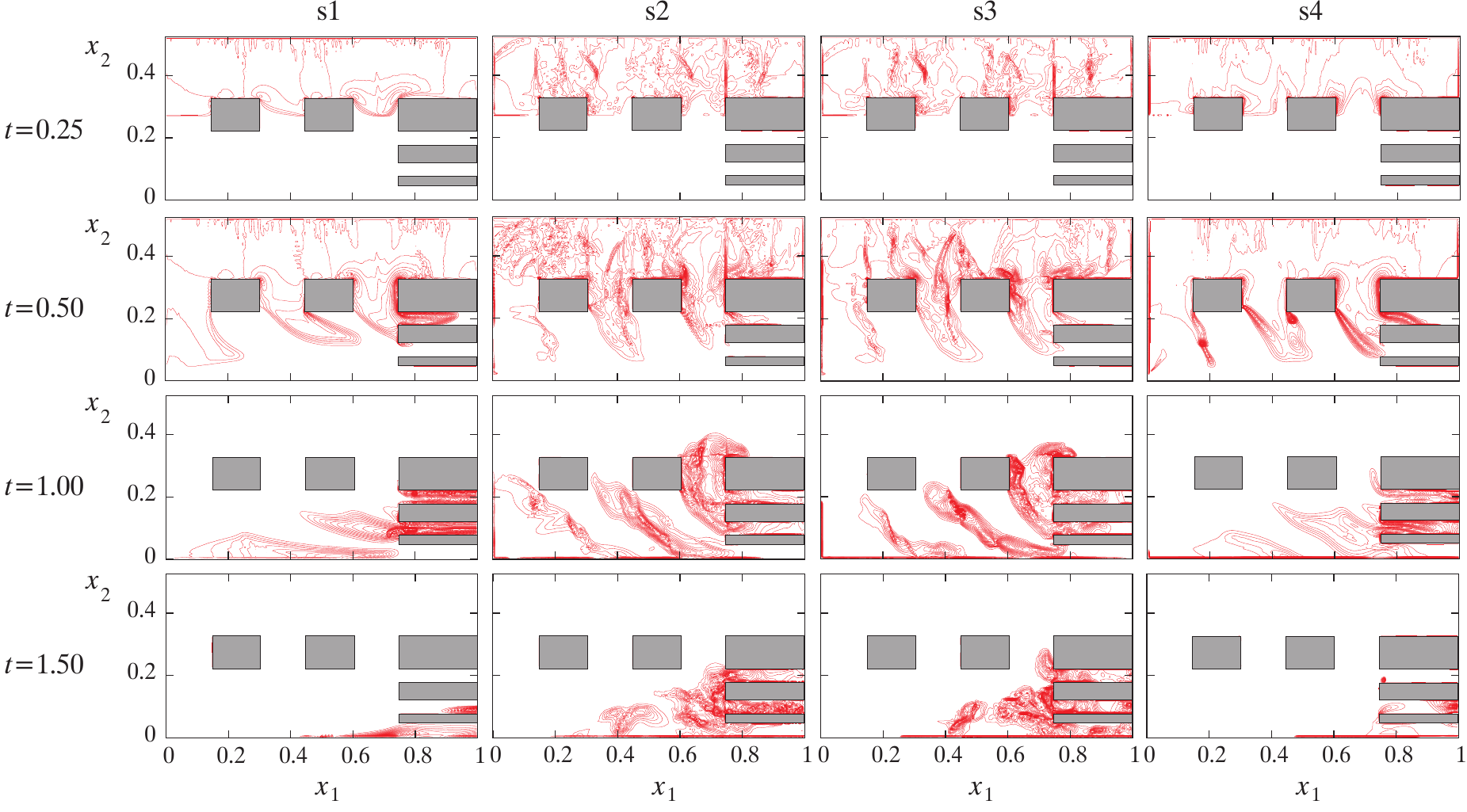}
\caption{Instantaneous crowd density fields (perception strategies column-wise, times row-wise)}
\label{fig.16}
\end{center}
\end{figure}

Let us comment on the crowd density spatio-temporal patterns resulting from the perception strategies s1, \dots, s4 (see Fig. \ref{fig.16}, showing $20$ contour levels equispaced in the range $[0.2,\,1]$).
\begin{itemize}
\item Strategy s1 is the most fluent: pedestrians do not hesitate to follow the desired direction because, the vectors $\e_d$, $\e_i$ being parallel (Eqs. \eqref{eq.ei}, \eqref{eq.xp.s1.2D}), it invariably results $\e_v=\e_d$ (Eq. \eqref{eq.dir}). Consequently, the distribution of the flow is essentially dictated by the domain geometry, chiefly by the arrangement of the obstacles in the walking area. Pedestrians getting through the right passage systematically head for the upper corridor, those getting through the middle passage head for the middle corridor, and finally those getting through the left passage head for the lower corridor. No significant swerves occur in front of the accesses of the corridors, therefore the total emptying time is the smallest one.

\item Strategies s2 and s3, quite similar to each other, are on the whole the less fluent. The crowd continuously disaggregates and aggregates because of the interactions, forming density lanes separated by low density areas. For instance, pedestrians getting through the right passage split off in two branches pointing towards the upper and the middle corridor, respectively, and leaving an almost empty space between them. Instead, pedestrians getting through the middle and the left passage ignore the upper corridor, preferring the middle and lower ones. This ultimately causes part of pedestrians coming from left and right passages to access the middle corridor. As a result, a higher crowding globally occurs, which slows down the total emptying of the station with respect to strategy s1. For reference, the estimated emptying time is approximately 1.5 times the one needed in strategy s1.

\item Strategy s4 is qualitatively intermediate between s1 and the pair s2, s3. Compared to the latter, the flow is more ordered and homogeneous: no sharp switches between high and low density zones are observed, because of the averaging implied by this perception strategy. However, the crowd does not simply follow the desired direction like in strategy s1. Indeed, pedestrians getting through the right passage split up again between the upper and the middle corridors, while those getting through the middle and the left passage head preferentially for the middle and the lower corridors. Nevertheless, the homogeneity of the flow at the access of these two corridors causes a relatively small crowding, which ultimately allows for a quicker emptying of the station with respect to strategies s2, s3. In this case, the emptying time is about 1.2 times the one needed in strategy s1.
\end{itemize}

\begin{figure}[!t]
\begin{center}
\includegraphics[width=0.6\textwidth,clip]{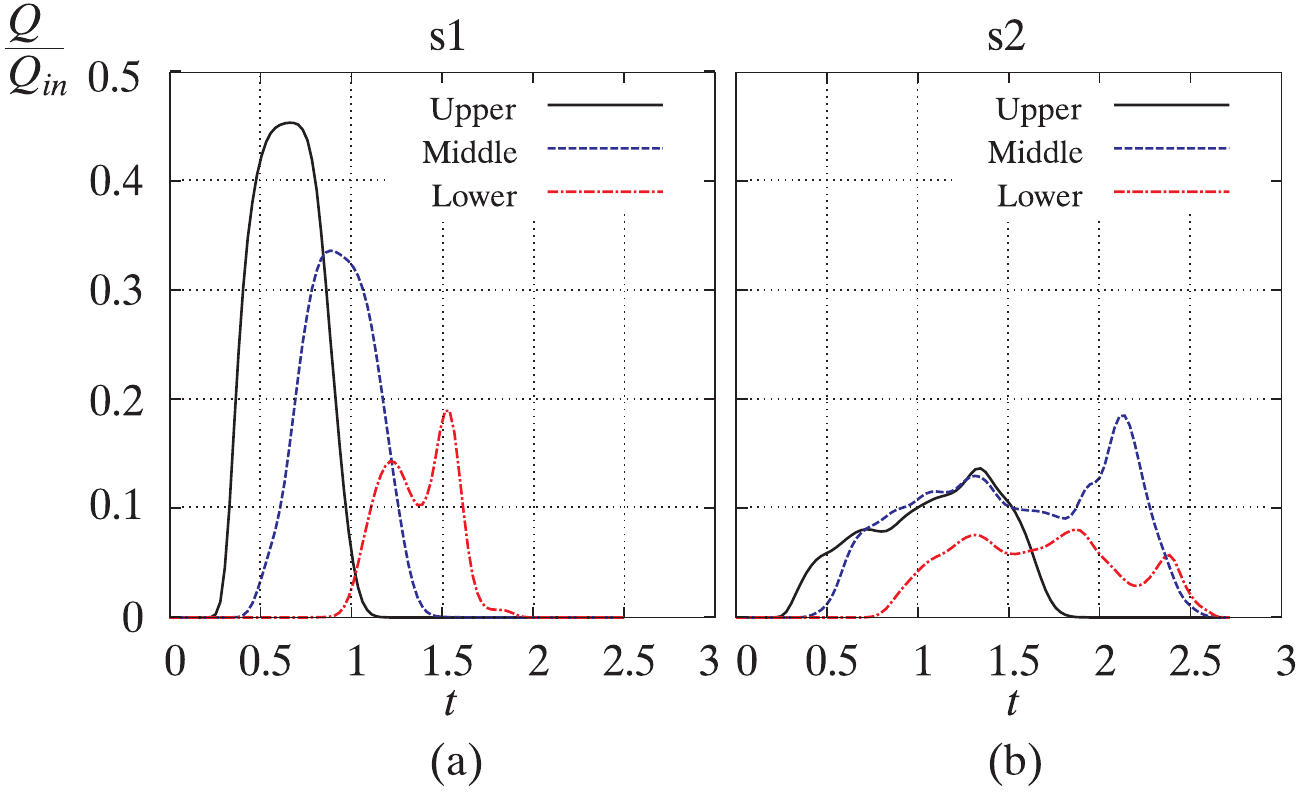} \\
\caption{Time evolution of the normalised integral fluxes $Q_i$ across the entrances of the three corridors: (a) strategy s1, (b) strategy s2}
\label{fig.17}
\end{center}
\end{figure}

In order to synthetically analyse the topological differences in the flows of strategies s1 and s2, the integral pedestrian flux $Q_i$ across the entrance $l_i$ of the $i$-th corridor is introduced: 
\begin{equation*}
	Q_i(t)=\lint_{l_i}\mathbf{q}(x_2,\,t)\cdot\mathbf{i}\,dx_2,
\end{equation*}
where $\mathbf{q}=\rho\vv$ is the vector-valued pointwise flux.
It is worth pointing out that each corridor width is $\vert l_i\vert=1$ m (see Fig. \ref{fig.15}a), thus the integral fluxes $Q_i$ are directly comparable. Figure \ref{fig.17} shows the time history of the fluxes, their values being normalised with respect to the (constant) integral flux imposed at the inlet boundary (corresponding to $0.22$ ped/s). Data coming from the numerical simulation have been processed by a low-pass filter in order to highlight the long-scale dynamics.

As far as strategy s1 is concerned, the following considerations can be proposed.
\begin{itemize}
\item The maximal fluxes in the corridors are attained sequentially in time, according to the increasing distance of each corridor from the platform. This is mainly due to that pedestrians keep the desired direction in the area from passages to corridors, hence the time needed to access the corridors is essentially a matter of distance to walk. Maxima are of different magnitudes, decreasing from the upper to the lower corridor, because the right passage serves the largest platform area (see Fig. \ref{fig.15}b) and, in addition, the flows coming from the three passages do not mix. In synthesis, the corridor fluxes in strategy s1 are confirmed to be completely driven by the desired direction and the passage locations.
\item The flux across the lower corridor features more than one local maximum, while the other fluxes are monotonically decreasing after a single maximum. This is caused by the delayed arrival of pedestrians getting through the left passage and walking along the bottom wall. The latter has a blockage effect on their velocity field, resulting in a reduced speed and an increased local density.
\end{itemize}

As far as strategy s2 is concerned, the fluxes in the three corridors are very different because of the crowd mixing induced by this behavioural strategy. The effect of such a mixing is visible in several features of the flux time history, that are outlined in the following.
\begin{itemize}
\item Local maximal fluxes have comparable magnitudes in all corridors, by far lower than those attained in the upper and middle corridor under strategy s1. Moreover, fluxes are significantly different from zero for a longer time interval than in s1. In other words, the mixing is responsible for a lower and more balanced exploitation of each corridor during a longer time.

\item The fluxes in the upper and the middle corridors reach their respective first local maximum at the same time $t\approx 1.3$ with the same magnitude. Indeed the flow getting through the right passage splits between the two corridors (see Fig. \ref{fig.16} at $t=0.5$ and $t=1$) as a consequence of the modelled pedestrian attitude to observe environmental conditions. Because of the same attitude, the flow getting through the middle passage swerves from the middle to the lower corridor and may induce the simultaneous local maximum in the latter flux (see Fig. \ref{fig.16} at $t=1.00$).

\item After merging in front of the corridors (see Fig. \ref{fig.16} at $t=1$ and $t=1.5$), the flows getting through the middle and the left passages head together for the lower corridor. Then the local density at the access of the latter increases, causing subsequently a new swerve of the flow from the lower to the middle corridor. This corresponds to the second local maximum of the flux in the middle corridor ($t\approx 2.2$) with simultaneous local minimum of the flux in the lower corridor.
\end{itemize}

On the basis of the results obtained from the simulations, strategy s2 seems the most appropriate for modelling pedestrian behaviour in an underground station under low density conditions, indeed it induces a realistic redistribution of the flow among all of the available corridors. In particular, from a design point of view it could be argued that the middle corridor is the most sought after by pedestrians, hence it is expected to play a key role in the emptying process of the station.

\section{Conclusions}
\label{four}
In this paper a phenomenological framework for the modelling of crowd dynamics has been proposed, based on the concept of behavioural strategy developed by walkers. Three main elements, leading to the definition of a behavioural strategy, have been identified: (i) the will to reach some \emph{destination}; (ii) a \emph{perceptional process}, which determines how pedestrians evaluate and synthesise the information coming from the environment; (iii) a \emph{decisional process}, which leads pedestrians to the final route choice based on the information elaborated in the previous two steps. In particular, the perceptional process introduces two crucial features of pedestrian behaviour, \emph{non-locality} and \emph{anisotropy}.

These ideas have been translated into a multidimensional mathematical model able to describe the macroscopic spatio-temporal patterns of pedestrian movement. The model is based on the mass conservation equation supplemented by a direct modelling of the velocity of the walkers, which incorporates the above-mentioned decision-based dynamics. In particular, the speed of the crowd is related to the macroscopic density perceived by pedestrians via a fundamental relation motivated by experimental data, whereas the direction of movement stems from the interactions among the individuals. 

The model has been used to address two specific real world applications: (i) crowd-structure interaction due to coupled dynamics of pedestrian flow and vibrating footbridge; (ii) pedestrian movement in a two-dimensional built environment (specifically, an underground station) with several obstacles and exits. \emph{Ad hoc} computational techniques have been developed, in order to conveniently face the difficulties posed by some non-standard terms of the equations (e.g., the non-local flux of the crowd) as well as by the coupling with the structure in 1D or by the interaction with the obstacles in 2D.

The results of the numerical simulations have shown that the behavioural strategy plays a fundamental role in determining the dynamics and the overall evolution of the system. This implies that a correct characterisation of the behavioural strategy is a major modelling task for obtaining useful qualitative and quantitative information from the model. It should be noticed that the strategy, and in particular the perceptional process, may change dramatically depending on the travel purpose and the emotional state of the crowd. In the paper a few alternative perception strategies have been proposed, ideally referring to resolute, anxious, and curious pedestrians. These may apply to different categories of walkers with different travel purposes, e.g., travellers catching a train, commuters in rush hours, shoppers.

Remarkably, panic has been out of the scope of the present work. The proper modelling of panic is probably not simply a matter of perception strategy, but involves a change also in the route choice and in the importance given to the desired direction of movement with respect to the interactions with other individuals. In \cite{bel4} it is suggested that pedestrians entering a panic state tend to follow chaotically other individuals, dropping any specific destination, therefore they are mostly attracted towards areas of high density rather than seeking the less congested paths. In addition, in \cite{helb_5,col3} some proposals for modelling the transition from regular to panic conditions in the framework of first-order models using speed-density laws are provided. All of these ideas may be duly included in the behavioural framework presented in this paper, and they certainly deserve further study also from the modelling point of view.

\section*{Acknowledgements}
The authors extend warm thanks to Claudio Canuto for the stimulating discussions concerning the numerical issues.

A. Tosin was funded by a post-doctoral research scholarship ``Compa\-gnia di San Paolo'' awarded by the National Institute for Advanced Mathematics ``F. Severi'' (INdAM).

\end{document}